\documentclass[twocolumn,floatfix,aps,prc,showpacs]{revtex4}
 
\usepackage[final]{epsfig}
\usepackage{amsmath}
\usepackage{bm}
\usepackage{graphicx}

\usepackage{color}

\begin{document}
 
\title{A systematic comparison of jet quenching in different fluid-dynamical 
models}
 
\author{Thorsten Renk}
\email{thorsten.i.renk@jyu.fi}
\author{Hannu Holopainen}
\email{hannu.l.holopainen@jyu.fi}
\affiliation{Department of Physics, P.O. Box 35, FI-40014 University of 
             Jyv\"askyl\"a, Finland}
\affiliation{Helsinki Institute of Physics, P.O. Box 64, FI-00014 University 
             of Helsinki, Finland}
\author{Ulrich Heinz}
\email{heinz@mps.ohio-state.edu}
\author{Chun Shen}
\email{shen@mps.ohio-state.edu}
\affiliation{Department of Physics, The Ohio State University, Columbus, 
             OH 43210, USA} 

\pacs{25.75.-q,25.75.Gz}

\begin{abstract}
Comparing four different (ideal and viscous) hydrodynamic models for the 
evolution of the medium created in $200\,A$\,GeV Au-Au collisions, combined 
with two different models for the path length dependence of parton energy 
loss, we study the effects of jet quenching on the emission-angle dependence 
of the nuclear suppression factor $R_{AA}(\phi)$ and the away-side per 
trigger yield $I_{AA}(\phi)$. Each hydrodynamic model was tuned to provide 
a reasonable description of the single-particle transverse momentum spectra 
for all collision centralities, and the energy loss models were adjusted 
to yield the same pion nuclear suppression factor in central Au-Au
collisions. We find that the experimentally measured in-plane vs. 
out-of-plane spread in $R_{AA}(\phi)$ is better reproduced by models 
that shift the weight of the parton energy loss to later times along its 
path. Among the models studied here, this is best achieved by energy loss
models that suppress energy loss at early times, combined with hydrodynamic
models that delay the dilution of the medium density due to hydrodynamic
expansion by viscous heating. We were unable to identify a clear tomographic
benefit of a measurement of $I_{AA}(\phi)$ over that of $R_{AA}(\phi)$. 
\end{abstract}
 
\date{\today}

\maketitle

\section{Introduction}
\label{sec1}

The expression 'jet tomography' is often used to describe the analysis of 
hard pQCD processes taking place inside the soft medium created in an 
ultrarelativistic heavy-ion collision, with the aim to study properties 
of the medium. In particular, the focus is often on the nuclear suppression 
of hard hadrons in A-A collisions compared with the scaled expectation 
from p-p collisions, due to loss of energy from the hard parton by 
interactions with the soft medium (see e.g. 
\cite{Tomo1,Tomo2,Tomo3,Tomo4,Tomo5}), expressed through the nuclear 
suppression factor $R_{AA}$. 

In comparing theoretical calculations with experimental data on $R_{AA}$, 
there are two main unknown properties of the medium: The nature of the 
parton-medium interaction, being closely connected with microscopic 
properties of the medium (such as the relevant degrees of freedom), and 
the evolution of the medium density distribution, being connected with 
macroscopic properties such as the thermodynamical parameters in a fluid 
description of the medium. While some attempts at systematic comparison 
of different models for the parton-medium interactions using the same 
fluid-dynamical model for the medium have been made in order to assess 
the effect of assumptions in the parton-medium interaction model 
\cite{SysJet1,SysJet2}, there is very little systematics available for 
the effect of different hydrodynamical models on jet quenching
observables other than the overall suppression ratio $R_{AA}(p_T)$ 
\cite{SysHyd1,SysHyd2}. 

What may have slowed the insight that there is a need to systematically 
understand the role of the medium density evolution is the fact that 
early comparisons with data were usually done on the basis of single-hadron 
suppression $R_{AA}$ for central collisions only, and it took some time 
before it was realized that this quantity is quite insensitive to model 
assumptions \cite{gamma-h}, especially when (as usually done) one model 
parameter governing the strength of the parton-medium interaction is fit 
to the data. The need for more differential observables, such as 
$R_{AA}(\phi)$ as a function of the angle of the observed hadron with 
the reaction plane for different centralities \cite{ReactionPlane} or 
the strength suppression $I_{AA}$ observed in hard back-to-back 
correlations \cite{SysHyd1}, to overcome this insensitivity was only 
realized later.

Such observables are primarily sensitive to the effective path length 
dependence of the energy loss. As one goes from central to peripheral 
collisions, both the mean density of the medium and the average 
path length needed for a hard parton to traverse the medium decrease. 
Within a given centrality class, $R_{AA}(\phi)$ is dominated by the 
change in path length, modulated by a weak directional dependence of the
average density probed by the parton. How precisely the path length and 
density change with centrality depends, however, on details of the 
hydrodynamical evolution.

The aim of this paper is to investigate in some detail the connection 
between high-$p_T$ observables and the bulk medium evolution. In 
particular, we try to identify those properties of a hydrodynamical 
model which have the strongest influence on high-$p_T$ observables. We 
do so by presenting a systematic study of the directional dependence of
the nuclear suppression factor $R_{AA}$ and the away-side yield in 
triggered back-to-back correlations, $I_{AA}$, for several parton-medium 
interaction models with different path length dependence and a number of 
different hydrodynamical models for the medium. The hope is to derive 
constraints for a combination of both medium evolution and parton-medium 
interaction models that can be used to eventually arrive at a detailed
understanding of the dynamics of ultrarelativistic heavy-ion collisions.

\section{Hydrodynamical models}
\label{sec2}

We describe the medium probed by the hard parton as a thermalized fluid. 
Its temperature and energy and particle densities evolve in space and 
time, due to hydrodynamic expansion driven by pressure gradients. In 
this work we use both ideal and viscous hydrodynamics to generate these
density profiles.

\subsection{Ideal hydrodynamics}
\label{sec2a}

In the ideal case we solve the hydrodynamic equations
\begin{equation}
\label{eq1}
   \partial_\mu T^{\mu\nu} = 0, \quad 
   \partial_\mu j^\mu_B = 0, \quad 
\end{equation}
where $T^{\mu\nu} = (\epsilon + P)u^\mu u^\nu - g^{\mu\nu}P$ is the 
stress-energy tensor, $j_B^\mu = n_B u^\mu$ is the baryon number current, 
$n_B$ is the net baryon number density, $\epsilon$ the energy density, and 
$P$ the pressure in the local rest frame which moves with fluid four-velocity 
$u^\mu$ in the global frame. The Equation of State (EoS) $P=P(\epsilon, n_B)$,
relating the pressure to the local energy and net baryon number density, 
closes the set of dynamical equations. 

For testing parton energy loss, we have at our disposal space-time profiles
of $\epsilon$, $P$ and temperature $T$ from two different ideal 
hydrodynamical models. The first of these \cite{Nonaka:2006yn} solves 
Eqs.~(\ref{eq1}) in 3+1 dimensions, propagating both $T^{\mu\nu}$ and
$j_B^\mu$. The second model \cite{HolRasEsk} simplifies the problem to 
2+1 dimensions by assuming longitudinal boost invariance (i.e. none of 
the physical quantities depend on space-time rapidity 
$\eta=\frac{1}{2}\ln[(t{+}z)/(t{-}z)]$) and setting the net baryon density everywhere
to zero (such that only the energy-momentum tensor $T^{\mu\nu}$ needs to
be evolved, using a simplified form $P(\epsilon)$ for the EoS). These 
approximations can be made since we are interested in energy loss only 
at mid-rapidity where, at RHIC energies, the net baryon density is very 
small. Both calculations use light-cone coordinates $(\tau, x, y, \eta)$, 
where $\tau = \sqrt{t^2{-}z^2}$ is the longitudinal proper time and 
$\eta$ is the space-time rapidity.

Both models use smooth energy density distributions as initial conditions, 
based on the densities of binary collisions and wounded nucleons 
\cite{Kolb:2001qz}. For details we refer to the original papers describing
the models \cite{Nonaka:2006yn,HolRasEsk}. The hydrodynamic evolution 
starts at initial time $\tau_0 = 0.6~(0.17)$\,fm/$c$ in the (3+1)-d 
((2+1)-d) model. The (3+1)-d model uses a bag model EoS with a first 
order phase transition at $T_c = 160$ MeV \cite{Nonaka:2006yn} whereas the
(2+1)-d model \cite{HolRasEsk} uses the EoS from Ref.~\cite{Laine:2006cp}. 
Both Equations of State assume chemical equilibrium among the hadrons in the
dilute resonance gas phase below $T_c$. Thermal hadron spectra are 
calculated using the conventional Cooper-Frye method \cite{Cooper}, where 
particle emission is calculated from a constant-temperature surface. The 
freeze-out temperature is $T_\mathrm{dec}=130~(160)$\,MeV for the (3+1)-d 
((2+1)-d) model. Strong and electromagnetic two- and three-particle decays 
of unstable hadrons are taken into account before comparing with experimental 
data.

\subsection{Viscous hydrodynamics}
\label{sec2b}

We also study parton energy loss in a medium whose space-time evolution
is computed from viscous hydrodynamics, by solving the second-order 
Israel-Stewart equations in 2+1 dimensions as described in 
Ref.~\cite{Song:2007fn}, assuming longitudinal boost invariance and zero 
net baryon density. Here the energy-momentum tensor of the fluid is 
decomposed as 
\begin{equation}
T^{\mu \nu} = (\epsilon{+}P) u^\mu u^\nu - P g^{\mu \nu} + \pi^{\mu \nu}
\label{eq2}
\end{equation}
which differs from the ideal fluid decomposition in Sec.~\ref{sec2a} 
by the appearance of the traceless and symmetric shear viscous pressure 
tensor $\pi^{\mu \nu}$ satisfying $u_\mu\pi^{\mu \nu}=0$. Effects from 
bulk viscosity are neglected as small compared to $\pi^{\mu\nu}$
\cite{Song:2009rh}. The energy-momentum conservation equations 
$\partial_\mu T^{\mu\nu}=0$ are supplemented by the Israel-Stewart 
\cite{Israel:1979wp,Muronga:2001zk} evolution equations for the viscous 
pressure components $\pi^{\mu\nu}$, see \cite{Song:2007fn} for details.

The viscous hydrodynamic energy density profiles studied here were obtained
with the Equation of State s95p-PCE described in \cite{SHHS} and
Appendix C of Ref.~\cite{Huovinen:2009yb}. It matches the latest lattice
QCD data of the EoS at high temperatures with a chemically frozen hadron 
resonance gas EoS at low temperatures \cite{Huovinen:2009yb,SHHS} that 
uses non-equilibrium chemical potentials \cite{Hirano:2002ds} to ensure 
preservation of the stable hadron ratios at their chemical freeze-out 
values as the system cools below the chemical decoupling temperature 
$T_\mathrm{chem}=165$\,MeV that has been experimentally established 
\cite{BraunMunzinger:2001ip}. 

In the viscous simulations we start the hydrodynamic evolution at
$\tau_0=0.4$\,fm/$c$ and decouple the hadron momentum spectra at 
$T_\mathrm{dec}=130$\,MeV. To compute the hadron spectra from the 
hydrodynamical output along the freeze-out surface we again use the 
Cooper-Frye prescription, but with a modified expression for the
distribution function, $f(x,p)=f_\mathrm{eq}(x,p)+\delta f(x,p)$,
where we add to the local equilibrium distribution a small viscous
correction $\delta f$ that depends on the viscous pressure components
$\pi^{\mu\nu}(x)$ at freeze-out and grows quadratically with $p_T$ (see
\cite{Song:2007fn,SHHS} for details). The specific shear viscosity is 
fixed at $\eta/s=0.2$, independent of temperature. 

In addition to the Glauber model initial conditions used in the ideal fluid 
dynamical models, we also study a set of viscous hydrodynamic evolution 
models based on Color Glass (CGC-fKLN) initial conditions 
\cite{Kharzeev:2000ph,Drescher:2006pi} which, for noncentral collisions, 
feature somewhat larger initial eccentricities and surface density 
gradients than the Glauber model profiles (see Fig.~1 in \cite{SHHS} 
for a comparison of these profiles) and thus generate more radial and 
elliptic flow. We will label viscous hydrodynamic simulations initiated
with Glauber model profiles as ``vGlb'', and those initiated with CGC-fKLN 
profiles as ``vCGC''. We will discuss the consequences of these differences 
on the directional dependence of energy loss suffered by a parton propagating 
through these fireballs.

Neither the ideal nor the viscous fluid simulations studied here account 
for event-by-event fluctuations of the initial shape and orientation of 
the collision fireball. Such calculations were recently reported both for
ideal \cite{Aguiar:2001ac,Holopainen:2010gz,Petersen:2010cw} and
viscous hydrodynamics \cite{Schenke:2010rr}. Source eccentricities and
anisotropic flow are most strongly affected by these fluctuations at
very small and very large impact parameters where the fireball is
either almost round or very small. Neither of these two situations is
of interest in our present study.

\subsection{Spectra and elliptic flow from the hydrodynamic models}
\label{sec2c}

Whereas the directional dependence of the soft hadron spectra (i.e.
their elliptic flow) reflects the momentum anisotropy of the hadron 
emitting source at freeze-out, the emission angle dependence of parton 
energy loss probes more directly the geometrical aspects of the fireball,
i.e. its spatial deformation. Still, it may matter whether the hard parton 
moves with or against the collective flow as it propagates through the
fireball, so the hydrodynamical models need to be tuned to give a reasonably
accurate representation of the momentum-space structure of the fireball,
as reflected in the final hadron spectra, before we test the influence 
of differences in their geometrical features on parton energy loss.

In this subsection we demonstrate that all models provide a reasonable 
description of the transverse momentum spectra of pions and protons over 
the entire range of collision centralities \footnote{Note that our goal 
  is {\em not} to achieve a perfect description of the measured hadron 
  spectra. It is unlikely \cite{SHHS} that such a description can be 
  obtained within a purely hydrodynamic approach; a fully realistic 
  dynamical model can not avoid describing the late hadronic freeze-out 
  stage microscopically with kinetic theory 
  \cite{Bass:2000ib,Hirano:2005xf}.}.  
Testing the spectra for both a very light and a heavy hadron species, 
which react differently to radial collective flow 
\cite{Schnedermann:1992ra}, ensures that the 
%
\begin{figure*}[hbt]
\includegraphics[width=0.49\linewidth,clip=]{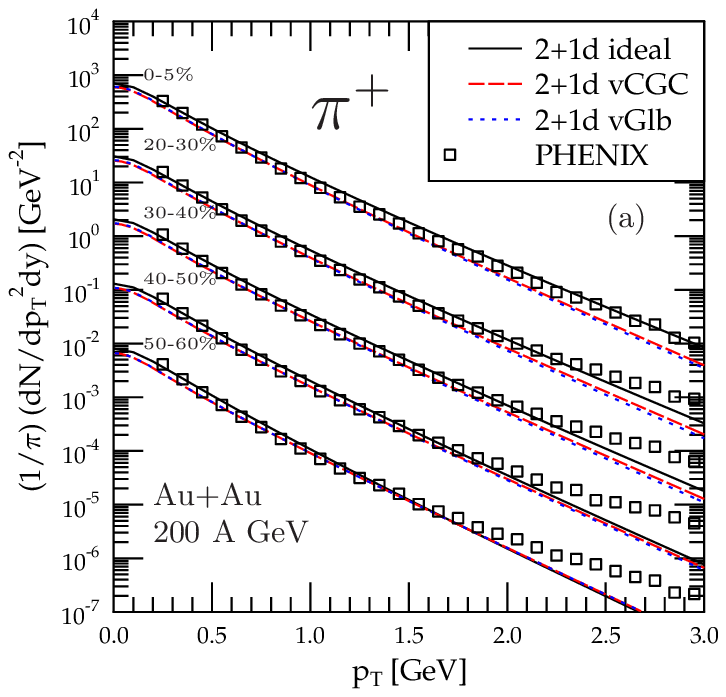}
\includegraphics[width=0.49\linewidth,clip=]{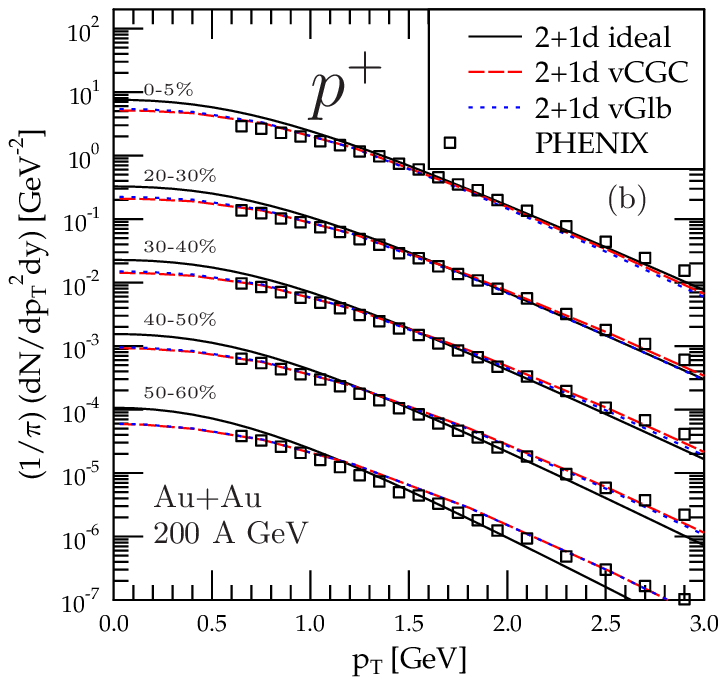}
\caption{\label{spectra} (Color online) Transverse momentum spectra of 
positively charged pions (a) and protons (b) from ideal and viscous 
(2+1)-d models for $200\,A$\,GeV Au-Au collisions at different centralities. 
Data from the PHENIX Collaboration \cite{Adler:2003cb} are shown without 
error bars since errors are smaller than the symbol size.}
\end{figure*}
%
appropriate amount of radial flow is generated in the evolution. The 
elliptic flow coefficient tests whether we also have the correct amount 
of flow anisotropy. Expressed in terms of velocity differences, the flow 
anisotropy is a small effect superimposed on a much larger radial flow
velocity; while this anisotropy is crucial in determining the transport
coefficients (in particular the shear viscosity) of the fireball fluid,
it is not expected to lead to major modifications of the directional 
dependence of parton energy loss. It affects the latter mostly by 
influencing the evolution of the spatial deformation of the fireball.
As we will see, the time dependences of the radius and shape of the fireball
play minor roles in the parton energy loss; hence, it is not a serious
problem for our analysis that the elliptic flow $v_2(p_T)$ is not well
reproduced by some of the hydrodynamical models we have studied.

\begin{figure*}[hbt]
\includegraphics[width=0.49\linewidth,clip=]{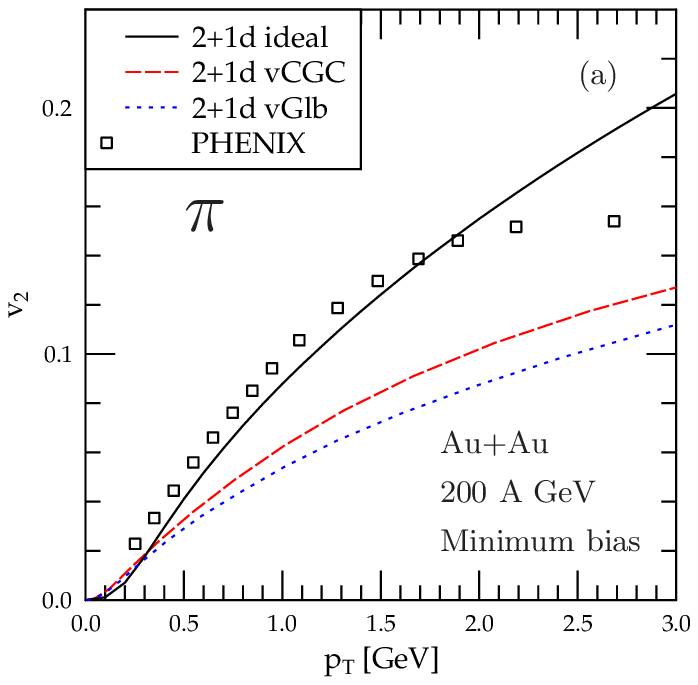}
\includegraphics[width=0.49\linewidth,clip=]{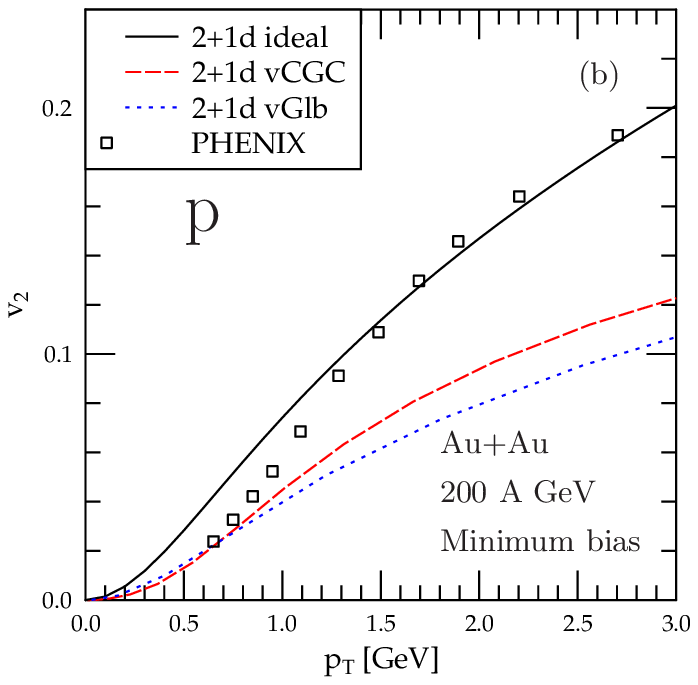}
\caption{\label{v2_mb} (Color online) Elliptic flow of (a) charged pions and 
(b) protons in minimun bias $200\,A$\,GeV Au-Au collisions. Data from the 
PHENIX Collaboration \cite{Adare:2006ti} are shown without error bars since 
errors are smaller than the symbol size.
}
\end{figure*}

\begin{figure*}[htb]
\includegraphics[width=0.49\linewidth,clip=]{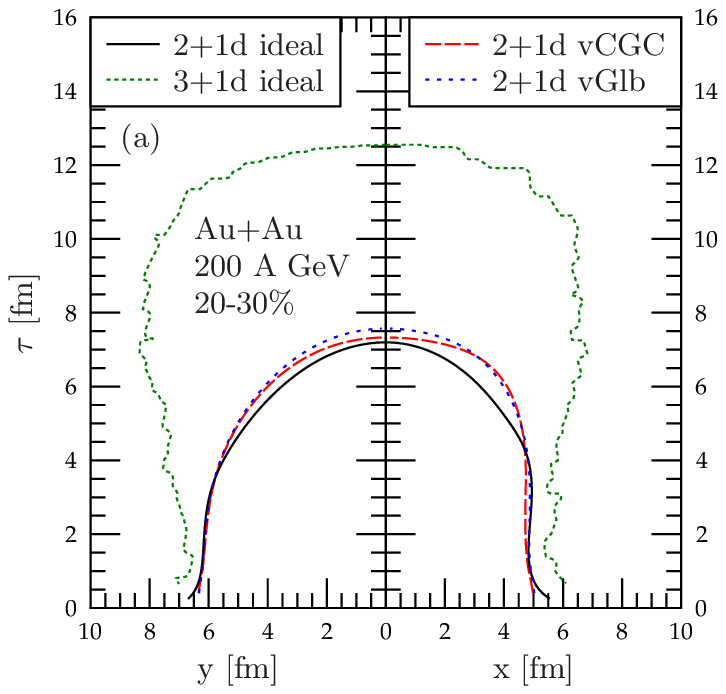}
\includegraphics[width=0.49\linewidth,clip=]{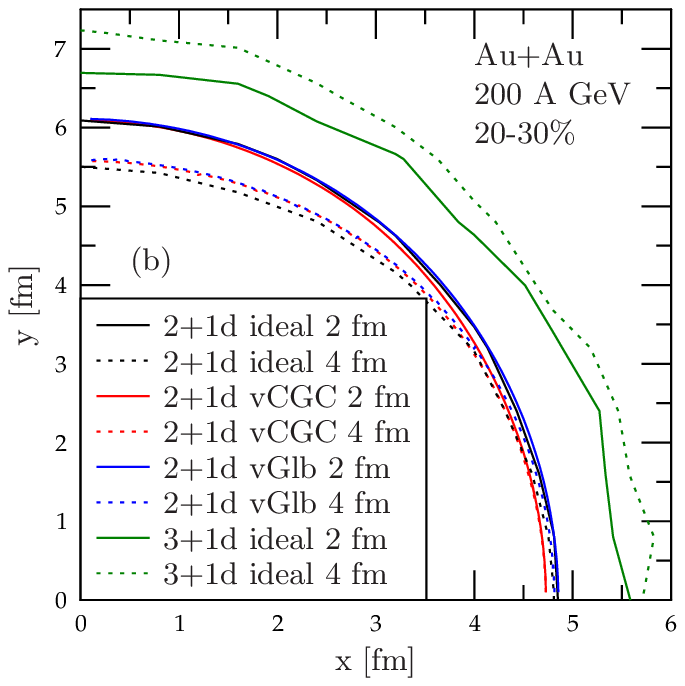}
\caption{\label{surfaces} (Color online) Freeze-out surfaces of the different 
hydrodynamic models in the (a) $r$-$\tau$ and (b) $x$-$y$ planes for 
$200\,A$\,GeV Au-Au collisions in the 20-30\% centrality class. The two 
halves of panel (a) show cuts through the freeze-out surface along the 
in-plane ($x$) and out-of-plane ($y$) directions, respectively. In panel (b),
solid (dashed) lines represent cuts at $\tau=2$\,fm/$c$ (4\,fm/$c$).
The contours for the three (2+1)-d hydrodynamic models are almost
indistinguishable.
}
\end{figure*}

Figure~\ref{spectra} shows the pion and proton spectra for $200\,A$\,GeV
Au-Au collisions at various centralities, as obtained from the three 
different (2+1)-d hydrodynamical models. Except for the peripheral bins, 
the viscous simulations give slightly steeper pion $p_T$-distributions 
than the ideal fluid ones. This is a consequence of the different chemical
composition in the hadronic phase as described in the preceding subsection 
\cite{Hirano:2002ds,Hirano:2005wx}. The proton spectra are markedly better
described by viscous than by ideal hydrodynamics. Shear viscosity adds to
the effective transverse pressure, generating more radial flow 
\cite{Song:2007fn} and thus making the spectrum harder, as desired by the
experimental data \footnote{In \cite{SHHS} it was shown that larger radial 
flow can also be generated by selecting lower freeze-out temperatures, with
similar effects on pion and proton $p_T$-spectra (but {\em not} their 
elliptic flow) as caused by larger viscosity (which matters most at early
times where the expansion rate is highest). This finding applies only,
however, if an equation of state with a chemically frozen hadron gas stage 
(such as s95p-PCE \cite{Huovinen:2009yb,SHHS}) is used \cite{Hirano:2005wx}; 
for chemically equilibrated equations of state, late radial flow makes not 
only the proton but also the pion spectrum harder which is inconsistent 
with the measured pion distribution.}. The somewhat steeper surface density 
gradients of the initial CGC energy density profile when compared to the 
Glauber model adds an additional small contribution to this effect.   
    
The $p_T$-dependent pion and proton elliptic flow from minimum bias
Au-Au collisions at $\sqrt{s}=200\,A$\,GeV is shown in Fig.~\ref{v2_mb}.
Here the viscous calculations are seen to badly underpredict the measured
values, as a consequence of shear viscous suppression of flow anisotropies
\cite{Song:2007fn,Romatschke:2007mq,Dusling:2007gi}. The disagreement
is not quite as bad for the CGC initial conditions which have somewhat
larger initial spatial eccentricity $\varepsilon_x=\frac
{\langle\!\langle y^2{-}x^2\rangle\!\rangle}
{\langle\!\langle y^2{+}x^2\rangle\!\rangle}$
\cite{Luzum:2008cw,Heinz:2009cv} where $\langle\!\langle\dots\rangle\!\rangle$
denotes an average over the energy density in the transverse plane.
But even in this case, the chosen value $\eta/s=0.2$ yields too much 
suppression of $v_2$, in agreement with the findings in \cite{Luzum:2008cw}.
We found in \cite{SHHS} that quite generally the measured flatness of the 
proton $p_T$-spectra and the large elliptic flow $v_2(p_T)$ are in tension 
with each other, the former preferring larger $\eta/s$ whereas the latter 
likes smaller viscosities. We here chose to optimize the slope of the 
proton spectra, i.e. the magnitude of radial flow. 

\subsection{Geometry of the hydrodynamic fireballs}
\label{sec2d}

The different initial and final conditions, equations of state and 
viscosities assumed in the different hydrodynamical fireball models
lead to some differences in the space-time evolution of the size
and shape of the fireball which, weighted with the corresponding
density distributions, affect the parton energy loss and its directional
dependence. Figure~\ref{surfaces} shows the freeze-out surfaces 
for Au-Au collisions in the 20-30\% centrality class ($b=7.49$\,fm)
in the $r$-$\tau$ and $x$-$y$ planes. The different shapes of the 
freeze-out contours along the $x$ and $y$ directions in 
Fig.~\ref{surfaces}(a) reflect the elliptical source deformation in
non-central collisions. Figure~\ref{surfaces}(b) shows freeze-out 
contours in the $x$-$y$ plane at times $\tau=2$\,fm/$c$ (solid) and 
$\tau=4$\,fm/$c$ (dashed), illustrating how the out-of-plane
elongation of the source decreases with time~\footnote{The freeze-out
surfaces extracted from the hydrodynamic output from the (3+1)-d simulations
are wiggly due to numerical effects related to the fact that these 
simulations were done in Lagrangian coordinates.}. 

The most striking feature of these contour plots is the similarity
of the freeze-out contours for the three (2+1)-dimensional models.
Even though the starting time $\tau_0$ for the viscous simulations 
(0.4\,fm/$c$) is more than twice as large as that used in the (2+1)-d 
ideal runs (0.17\,fm/$c$), and the viscous fluid is allowed to cool
down to $T_\mathrm{dec}=130$\,MeV (compared to $T_\mathrm{dec}=160$\,MeV
for the (2+1)-d ideal runs), all three models complete their freeze-out 
almost at the same time ($\approx 7.2-7.5$\,fm/$c$ for $b=7.49$\,fm). 
This is due to the additional radial flow produced by shear viscous 
pressure and the use in the viscous flow simulations of EoS s95p-PCE  
which is stiffer around the phase transition than the EoS used in
the ideal fluid simulations \cite{SHHS}. The combination of these two 
effects allows the viscous fireball to cool faster during the late evolution 
stages and thus freeze out sooner than an ideal fluid with the same 
initial conditions \cite{Song:2007fn}. 

During the early stages, on the other hand, viscous heating delays the 
cooling process, so at early times (up to about 4\,fm/$c$) the fireball 
center remains hotter and denser in the viscous case than for the ideal 
fluid \footnote{This information is not contained in Fig.~\ref{surfaces} 
but can be verified by consulting Ref. \cite{Song:2007fn}.}. The somewhat 
steeper initial density gradients of the CGC-fKLN profile (see Fig.~1 in 
\cite{SHHS}) generate slightly larger radial flow in the vCGC model, 
causing it to fully decouple a fraction of a fm/$c$ earlier than the 
vGlb model.
  
Compared to the (2+1)-d simulations, the space-time volume covered by
the ideal (3+1)-d fluid is much larger (wiggly green lines in 
Fig.~\ref{surfaces}). This not a consequence of dramatically different
transverse expansion in (3+1)-d and boost-invariant (2+1)-d evolution
(near midrapidity the longitudinal density profiles and expansion 
velocities in the (2+1)-d and (3+1)-d simulations are very similar), 
nor is it primarily due to starting the (3+1)-d simulation later (at 
$\tau_0=0.6$\,fm/$c$). The main reason is that the (3+1)-d simulations
use a bag model EoS with a first order phase transition that produces 
a relatively long-lived mixed phase where the speed of sound vanishes and 
the fluid stops accelerating.  

\section{Parton-medium interaction models}
\label{sec3}

In this work, we use two different models for the parton-medium interaction. 
The first is the Armesto-Salgado-Wiedemann (ASW) model of medium-induced 
radiative energy loss in perturbative Quantum Chromodynamics (pQCD), in 
the formulation of energy-loss probability distributions, so-called 
'quenching weights' \cite{QuenchingWeights}. As characteristic for 
perturbative models of medium-induced radiation, it leads to a quadratic 
dependence of mean energy loss with the in-medium path length $L$ in a 
constant medium due to the LPM suppression of subsequent radiation 
processes. We have picked this model among other formulations of 
radiative energy loss, since it shows the strongest path length dependence 
for $R_{AA}(\phi)$ of all radiative energy loss models tested in the same 
hydrodynamical background \cite{SysJet1}.

The second model is based on strong-coupling ideas for the medium. It 
is a hybrid model, in which the hard scales in the process are treated 
perturbatively, as in the standard pQCD radiative energy loss calculations, 
while the interaction with the plasma which involves strong-coupling 
dynamics is modeled by AdS/CFT calculations for the $N{\,=\,}4$ 
super-Yang-Mills (SYM) theory \cite{StrongCoupling}. In a constant medium, 
this approach leads to an $L^3$ dependence of mean energy loss. In the 
following, we will refer to this approach by the label 'AdS'.

We have refrained from testing a third class of models of energy loss 
by elastic scattering of a hard parton with medium constituents, which, 
as such scatterings are incoherent, would result in a linear dependence 
on the path length $L$ in a constant medium. However, a large contribution 
of such processes to the total energy loss can be ruled out already by 
the data for $R_{AA}(\phi)$ \cite{Elastic1,Elastic2}.

A key quantity in both models is the quenching weight, i.e. the energy loss 
probability distribution $P(\Delta E)$ given the path of a parton through 
the medium. In both models this is obtained by calculating the integrated 
virtuality transfer from the medium to the hard parton $Q_s^2$ and the 
characteristic medium-induced gluon energy $\omega_c$ by line integrals 
along the hard parton trajectory through the medium. Making use of a 
scaling law \cite{Scaling}, we use $Q_s$ and $\omega_c$ with the numerical 
results of \cite{QuenchingWeights} to obtain $P(\Delta E | \omega_c, Q_s) 
\equiv P(\Delta E)_\mathrm{path}$. What is different between the two models 
is the argument of the line integrals.

In the ASW model the medium is characterized by a transport coefficient 
$\hat{q}$ which measures the ability of the medium to transfer virtuality 
per unit path length. We assume that this can be written as a function of 
the medium thermodynamic parameters and parton position along the 
trajectory $\xi$ via the relation
\begin{equation}
\label{E-qhat}
  \hat{q}(\xi) = K \cdot 2 \cdot \epsilon^{3/4}(\xi) 
                 \bigl(\cosh \rho(\xi) - \sinh \rho(\xi) \cos\alpha(\xi)\bigr)
\end{equation}
between the local transport coefficient $\hat{q}(\xi)$ (specifying 
the quenching power of the medium), the energy density $\epsilon$ and 
the local flow rapidity $\rho$ with angle $\alpha$ between flow and 
parton trajectory \cite{Flow1,Flow2}. We view the parameter $K$ as a 
tool to account for the uncertainty in the selection of the strong 
coupling $\alpha_s$ and possible non-perturbative effects increasing 
the quenching power of the medium (see discussion in \cite{Correlations}) 
and adjust it such that the pionic $R_{AA}$ measured in central Au-Au 
collisions is reproduced.

With this expression for $\hat{q}$, we evaluate for each path through the 
medium (given by the initial vertex position $\bm{r}_0 = (x_0, y_0)$ in 
the transverse plane and the angle $\phi$ of the outgoing parton with the 
reaction plane) the line integrals
\begin{equation}
\label{E-lineintegral1}
  Q_s(\bm{r}_0,\phi) \equiv \langle \hat{q} L \rangle 
                          = \int d\xi\, \hat{q}(\xi) 
\end{equation}
and
\begin{equation}
\label{E-lineintegral2}
  \omega_c (\bm{r}_0,\phi)= \int d\xi\, \xi\, \hat{q}(\xi).
\end{equation}
In the AdS model, these expressions are changed into \cite{StrongCoupling}
\begin{equation}
\label{E-lineintegral3}
  Q_s(\bm{r}_0,\phi) = K \int d\xi\, \xi\, T^4(\xi) 
\end{equation}
and
\begin{equation}
\label{E-lineintegral4}
  \omega_c(\bm{r}_0,\phi) = K \int d\xi\, \xi^2\, T^4(\xi),
\end{equation}
where $T(\xi)$ the local temperature of the medium and $K$ is again a 
(different) free parameter, to be adjusted such that the pionic $R_{AA}$ 
for central Au-Au collisions is reproduced.

From the energy loss distribution for a given single path we can define 
the averaged energy loss probability distribution for a given angle $\phi$ 
as
\begin{equation}
\label{E-P_phi}
  \langle P(\Delta E)\rangle_\phi \negthickspace = 
  \int_{-\infty}^{\infty} \negthickspace \negthickspace \negthickspace 
  \negthickspace dx_0 
  \int_{-\infty}^{\infty} \negthickspace \negthickspace \negthickspace 
  \negthickspace dy_0\, P(x_0,y_0)\, P(\Delta E)_\mathrm{path},
\end{equation}
where for given impact parameter $\bm{b}$ the probability density of hard 
vertices in the transverse plane $P(x_0,y_0)$ is given by the product of 
the nuclear profile functions as
\begin{equation}
\label{E-Profile}
  P(x_0,y_0) = \frac{T_{A}(\bm{r}_0{+}\bm{b}/2) T_A(\bm{r}_0{-}\bm{b}/2)}
                    {T_{AA}(\bm{b})},
\end{equation}
and the nuclear thickness function is given in terms of the Woods-Saxon 
nuclear density $\rho_{A}(\bm{r},z)$ as 
\begin{equation}
  T_{A}(\bm{r}) = \int dz\, \rho_{A}(\bm{r},z).
\end{equation}

We calculate the momentum spectrum of hard partons in leading order 
perturbative QCD (LO pQCD) (explicit expressions are given in 
\cite{Correlations} and references therein). The medium-modified 
perturbative production of hadrons at angle $\phi$ can then be 
computed from the expression
\begin{equation}
\label{E-LOpQCD}
  \frac{d\sigma_{med}^{AA\rightarrow h+X}}{d\phi} \negthickspace 
  \negthickspace 
  = \sum_f \frac{d\sigma_{vac}^{AA \rightarrow f +X}}{d\phi} 
    \otimes \langle P(\Delta E)\rangle_\phi 
    \otimes D_{f \rightarrow h}^{vac}(z, \mu_F^2),
\end{equation} 
with $D_{f \rightarrow h}^{vac}(z, \mu_F^2)$ the fragmentation function 
with momentum fraction $z$ at scale $\mu_F^2$ \cite{KKP}. From this we 
compute the nuclear modification function $R_{AA}$ vs. reaction plane as
\begin{equation}
  R_{AA}(P_T,y,\phi) = \frac{dN^h_{AA}/dP_Tdy d\phi}
                            {T_{AA}(\bm{b})\,d\sigma^{pp}/dP_Tdy d\phi}.
\end{equation}

The suppression of back-to-back high-$p_T$ hadron correlations is computed 
in a Monte-Carlo (MC) framework \cite{I_AA_RP}. We start from the 
expression for the production of two hard partons $k,l$ in LO pQCD which 
is described by
\begin{equation}
\label{E-2Parton}
  \frac{d\sigma^{AB\rightarrow kl +X}}{dp_T^2 dy_1 dy_2} \negthickspace 
  = \sum_{ij} x_1 f_{i/A}(x_1, Q^2) x_2 f_{j/B} (x_2,Q^2) 
    \frac{d\hat{\sigma}^{ij\rightarrow kl}}{d\hat{t}}
\end{equation}
where $A$ and $B$ stand for the colliding objects (protons or nuclei) and 
$y_{1(2)}$ is the rapidity of parton $k(l)$. The distribution function of 
a parton type $i$ in $A$ at a momentum fraction $x_1$ and a factorization 
scale $Q \sim p_T$ is $f_{i/A}(x_1, Q^2)$. The distribution functions are 
different for free protons \cite{CTEQ1,CTEQ2} and nucleons in nuclei 
\cite{NPDF,EKS98}. The fractional momenta of the colliding partons $i$, 
$j$ are given by $ x_{1,2} = \frac{p_T}{\sqrt{s}} \left(\exp[\pm y_1] 
+ \exp[\pm y_2] \right)$.

By sampling this expression, we generate events of back-to-back parton 
pairs which are placed on a vertex sampled according to Eq.~(\ref{E-Profile}) 
for given orientation $\phi$ with respect to the reaction plane. Given 
$(\bm{r}_0,\phi)$, we compute $P(\Delta E)$ for both partons according 
to the procedure outlined above and sample the distribution to obtain the 
energy loss for the given event.

Finally, we convert the simulated partons into hadrons. Note that this 
cannot be done using a fragmentation function as in Eq.~(\ref{E-LOpQCD}) 
since $D_{f \rightarrow h}^{vac}(z, \mu_F^2)$ takes a hadronic energy 
scale $\mu_F$ as argument and measures the inclusive hadron yield, whereas 
we are interested in the yield of leading hadrons given a partonic energy 
scale. 

More precisely, in 
order to determine if there is a trigger hadron above a given threshold, 
given a parton $k$ with momentum $p_T$, we need to sample 
$A_1^{k\rightarrow h}(z_1, p_T)$, i.e. the probability distribution to 
find a hadron $h$ from the parton $k$ where $h$ is the most energetic 
hadron of the shower and carries the momentum $P_T = z_1 \cdot p_T$. In 
the following, we make the assumption that the hadronization process itself, 
at least for the leading hadrons of a shower, happens well outside the 
medium. As a consequence, we neglect any interaction of formed hadrons 
with the medium. The time scale for hadronization of a hadron $h$ in its 
rest frame can be estimated by the inverse hadron mass, $\tau_h \sim 1/m_h$; 
boosting this expression to the lab frame one finds $\tau_h \sim E_h/m_h^2$. 
Inserting a hard scale of 6 GeV or more for the hadron energy and the pion 
mass in the denominator (as pions constitute the bulk of hadron production), 
this assumption seems well justified. We extract $A_1(z_1, p_T)$ and the 
conditional probability $A_2(z_1, z_2, p_T)$ to find the second most 
energetic hadron at momentum fraction $z_2$ {\em given that the most 
energetic hadron was found with fraction $z_1$} from  HERWIG \cite{HERWIG}. 
After hadronization, we check if the most energetic hadron fulfills a 
given trigger condition and, if yes, we count the yield in various 
momentum bins of hadrons back-to-back with the trigger. Finally, we 
obtain the suppression factor $I_{AA}(\phi)$ for given trigger and 
associate momentum windows by dividing by the per-trigger yields found 
with nucleon parton distributions \cite{CTEQ1,CTEQ2} in the absence of 
a medium. The procedure is described in detail in \cite{I_AA_RP}.

\section{The bulk fluid medium evolution 'seen' through hard probes}
\label{sec4}

Hard partons undergoing energy loss do not probe the same properties of 
the bulk medium as soft hadrons, or they probe them in a different way. 
For example, while the coefficient $v_2$ in the soft sector measures 
pressure gradients translating an initial spatial anisotropy in 
non-central collisions into a momentum-space anisotropy, the same 
coefficient for high $P_T$ hadrons measures directly the spatial 
anisotropy through the different energy loss induced by different 
densities seen by hard partons as a function of their angle with the 
reaction plane. This difference in the underlying physics is the reason 
why we prefer to present and discuss our results in terms of 
$R_{AA}(\phi)$ rather than in terms of the mean $R_{AA}$ and $v_2$ at 
high $P_T$ for a centrality class. While both choices contain the same 
information, we feel that $R_{AA}(\phi)$ emphasizes the underlying 
suppression process.

To give a second example, while $m_T$-spectra of soft hadrons are 
rather sensitive to the late-time hadronic evolution of the medium 
and the amount of flow created during the hadronic evolution, the 
medium modification of hard probes is not at all sensitive to late 
time dynamics. The reason is that hard partons propagate through the 
medium with the speed of light, and thus typically escape from the 
medium at time scales of order of the size of the overlap region. This 
is especially true for observed hadrons, which have a bias to be 
produced relatively close to the surface \cite{Correlations}.

In the following, we discuss some features of hydrodynamical models 
that are likely candidates to be probed by energy loss.

\subsection{Medium properties potentially probed by energy loss}
\label{sec4a}

As apparent from Eqs.~(\ref{E-lineintegral1}) to (\ref{E-lineintegral4}), 
the medium is probed by the energy loss models through path-length-weighted 
line integrals over the medium energy density $\epsilon$ or temperature 
$T$. The value of these integrals thus depends on
\begin{itemize}
\item the lower limit of the integral, corresponding to the time at which 
secondary particle production starts to be important enough to induce 
energy loss. Usually, the equilibration time $\tau_0$ of the hydrodynamical 
model is used here, but there is no reason in principle why a
non-equilibrated medium, if sufficiently dense, could not induce energy 
loss. However, the sensitivity to the lower limit is expected to be 
comparatively weak, as factors of $\xi$ or $\xi^2$ in the $\omega_c$ 
integrals suppress this region. In a Bjorken model, $\hat{q}$ would 
diverge as $1/\xi$ for small times, thus cancelling a factor $\xi$ of 
the suppression. However, prior to equilibration the density of particles 
in the medium off which the hard parton can scatter, thereby inducing it 
to radiate gluons, must generically be smaller than in thermal equilibrium
(which maximizes entropy and is thus a state of maximum particle density for
given energy density), so this cancellation cannot be perfect, and we expect 
(within reasonable limits) a weak sensitivity to the choice of the initial 
time.

\item the upper limit of the integral, corresponding to the time scale 
at which the hard parton is no longer surrounded by a medium. This is 
usually assumed to be given by the location of the Cooper-Frye surface 
in a hydrodynamical model beyond which the medium is no longer coupled 
but free-streaming. The Cooper-Frye surface in turn is often defined to 
be an isothermal surface. Since the Cooper-Frye prescription is clearly 
an idealization of the real physics of the system boundary, there is 
again no strong reason to identify the upper integration boundary with 
this surface, as there could be energy loss of a parton in a weakly coupled 
hadronic halo. However, again in practice the sensitivity to the detailed 
choice of the parameter is parametrically weak. While the factors $\xi$ 
or $\xi^2$ tend to enhance late-time contributions to the integral, the 
medium density at late times or large distances from the center dilutes 
eventually like $1/\xi^3$, due to both longitudinal and transverse flow. 
Thus, beyond a point we do not expect that our results depend strongly 
on the medium boundary choice.

\item the functional form of the integrand itself. Note that what is 
probed by the parton is not the medium density at any given proper time, 
but rather the medium density along the light cone. However, since 
interference effects suppress early time energy loss and the onset of 
transverse flow suppresses late time energy loss, effectively the mean 
energy loss per unit time $dE/dx$ reaches a relatively sharp maximum 
around $\tau_\mathrm{peak} \sim 3-4$ fm/c (cf. \cite{Mach1}). Thus, to first approximation, 
energy loss probes the density distribution of the medium around 
$\tau_\mathrm{peak}$. In particular, this implies that there is a finite 
time for processes like viscous entropy production or decay of the 
spatial deformation by pressure-generated anisotropic flow to modify 
the state of the system before it is probed by energy loss. 

\item finally, when comparing the distribution (\ref{E-Profile}) of primary
production vertices with any transverse distribution of matter in a 
hydrodynamical model, one will find that there is always a non-zero 
probability to find a hard vertex outside the medium and hence partons 
which never experience energy loss. Since such 'halo-partons' never probe 
the medium, they show no correlation with the reaction plane angle $\phi$ 
and hence their presence will dilute any dependence of $R_{AA}$ on $\phi$. 
This effect scales with the size of the hydrodynamical medium, i.e. with 
the assumed initial extent of the Cooper-Frye surface. For very large media
this 'halo effect' is suppressed, but for drastic assumptions (such as 
energy loss only in the QGP) it is not negligible.
\end{itemize}

In the following, we will discuss these points in more detail in the 
context of different hydrodynamical models.

\subsection{Viscous entropy production}
\label{sec4b}

In both ideal and viscous fluid dynamics, the entropy of the final state is 
fixed by the observed multiplicity and chemical composition of the emitted
hadrons. But only in ideal fluid dynamics this final entropy agrees with its
initial value. In viscous hydrodynamics, viscous heating causes the entropy 
to increase with time, which implies that, for the same final multiplicity,
the viscous fluid starts at lower entropy and parton density than the ideal
fluid. The rate of viscous entropy production grows with the expansion rate
of the fireball. A boost-invariant longitudinal expansion profile, as 
initially assumed in all versions of the hydrodyamic model studied here,
leads to an expansion rate that diverges like $1/\tau$ at early times $\tau$.
Thus, most of the viscous entropy production happens near the beginning of
the expansion \cite{Song:2007fn}. Viscous heating causes the fireball
density to decrease initially more slowly with time than for an ideal fluid.
This has implications for the density profile seen by a hard parton 
propagating on the lightcone. 

\begin{figure}[htb]
\epsfig{file=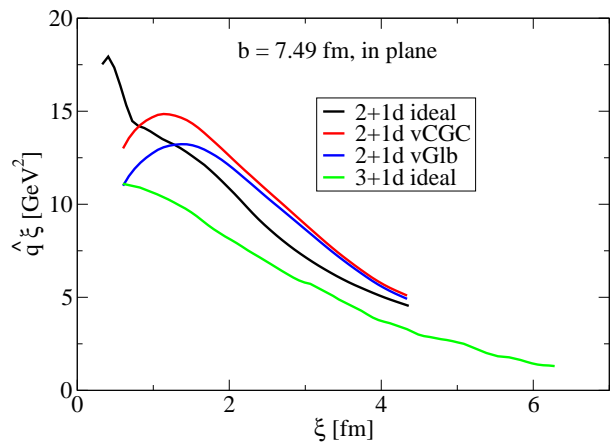,width=8cm}
\caption{\label{F-viscous-entropy} (Color online) The integrand of 
Eq.(\ref{E-lineintegral2}) as a function of parton distance $\xi$ along 
the light-cone for four different hydrodynamical models, shown for a 
parton propagating in-plane from the fireball center.
}
\end{figure}

We illustrate this effect in Fig.~\ref{F-viscous-entropy} where we show
the integrand of Eq.~(\ref{E-lineintegral2}), $\xi\,\hat{q}(\xi)$, as a 
function of $\xi$ for a parton  propagating in-plane from the medium 
center in a hydrodynamical background corresponding to Au-Au collisions 
at impact parameter $b = 7.49$\,fm, for four different hydrodynamical models. 

While the ideal hydrodynamical models show a monotonical decrease
for $\xi\,\hat{q}(\xi)$, the viscous models show an initial rise 
due to viscous entropy production, followed by a decrease due to 
flow-driven density dilution. This means that in viscous hydrodynamical 
models energy loss is generically somewhat shifted to later times. 

Note also that due to the lower entropy in the initial state in viscous 
hydrodynamics, if one wants to get the same $R_{AA}$, the factor $K$ in 
Eqs.~(\ref{E-qhat}), (\ref{E-lineintegral3}), (\ref{E-lineintegral4}) 
must be larger (by about a factor 2 for the models presented here) than 
in an ideal fluid model that leads to a similar final state.

\subsection{Line integral limits}
\label{sec4c}

As discussed above, there is no compelling reason why the hydrodynamical 
thermalization time should be equal to the initial time of energy loss, 
nor why the Cooper-Frye decoupling surface for soft hadrons should 
coincide with the boundary beyond which the hard parton no longer 
interacts with the medium.

In order to test the sensitivity of our results to choices of these 
parameters different from the hydrodynamical values, without explicitly 
modeling the (rather complicated) dynamics prior to thermalization or 
the perturbative scattering of hard quarks or gluons with hadrons in the 
halo, we adopt the following procedure: We probe the response of the 
system to variations of the hydrodynamical starting time $\tau_0$ and
of the final temperature $T_\mathrm{F}$, with the understanding that 
these parameters are not meant as physically reasonable choices within 
the hydrodynamical framework, but rather serve to generate upper limits 
for the true conditions --- 
certainly prior to thermalization the actual energy density will be less 
than in the hydrodynamical extrapolation to early times. In that sense, 
the parameters $\tau_0$ and $T_\mathrm{F}$ are to be understood in this 
subsection as the initial energy loss time and the equivalent temperature 
corresponding to the edge of the medium that contributes to energy loss.

\begin{figure}[htb]
\epsfig{file=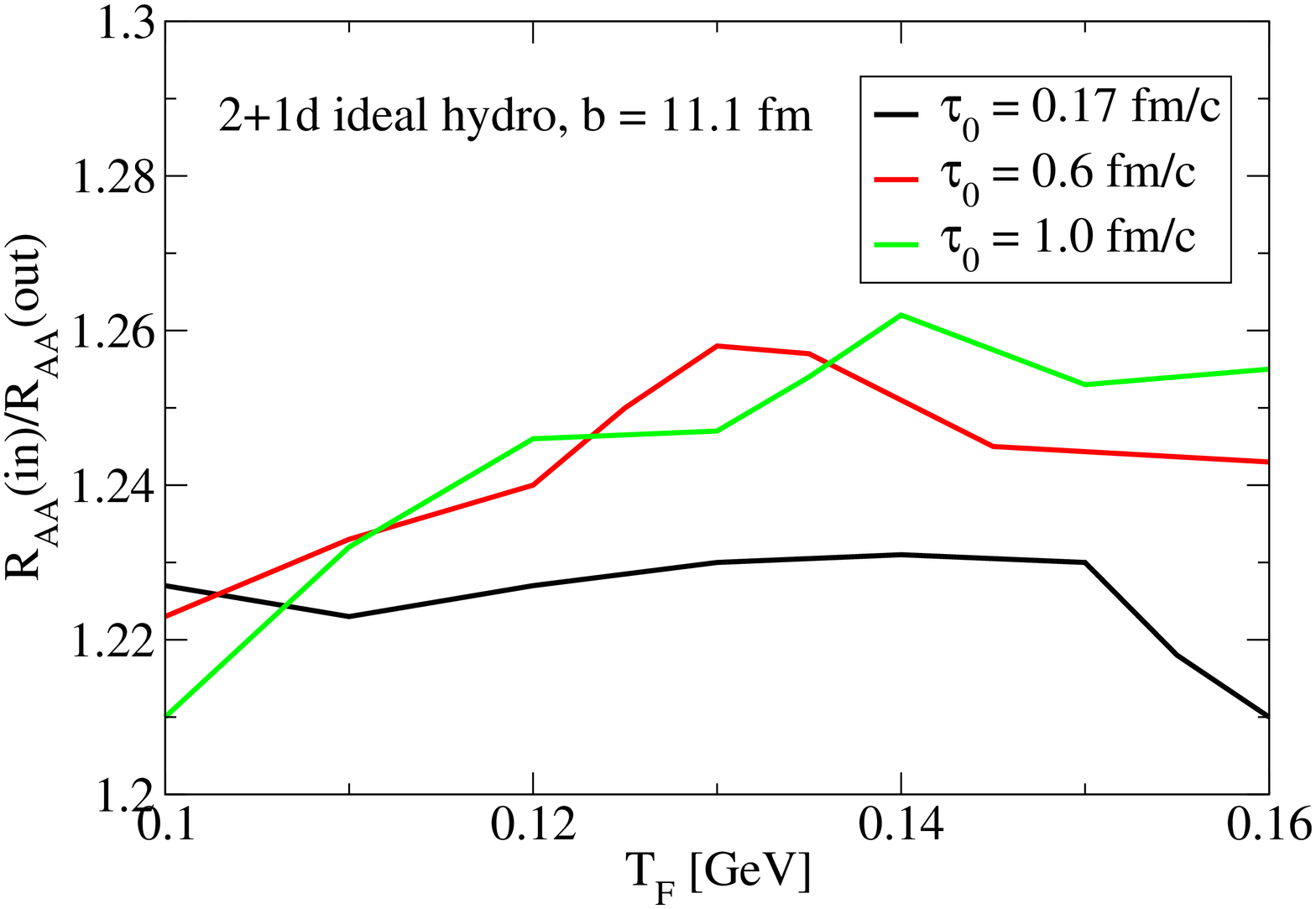, width=8cm}
\caption{\label{F-line-integrals} (Color online) The ratio of $R_{AA}(\phi{\,=\,}0)/R_{AA}
\left(\phi{\,=\,}\frac{\pi}{2}\right)$ for various combinations of the 
initial energy loss time $\tau_0$ and the equivalent temperature 
$T_\mathrm{F}$ of the medium edge, for the case of the (2+1)-d ideal fluid 
model.
}
\end{figure}

In Fig.~\ref{F-line-integrals} we show the results of a variation of 
$\tau_0$ and $T_\mathrm{F}$ on the observable spread between the in-plane 
and out-of-plane emission, for the case of the (2+1)-d ideal hydrodynamical 
model. While the variation looks optically significant, it is small on an 
absolute scale (note the suppressed zero!), at most 15\% from the mean.
This is less than, for example, the variation between different 
hydrodynamical models. This in itself is certainly reassuring, as it 
quantifies the uncertainty in choosing the proper line integral limits. It 
is also readily apparent that the spread is typically maximized for 
'reasonable' choices of the freeze-out temperature, thus there is no 
evidence for a need to choose dramatically different last-scattering 
surfaces for soft and hard particles.

Note that there is a systematic trend that large $\tau_0$ (i.e. delayed
parton energy loss) leads to an increased in-plane vs. out-of-plane spread. 
This fits well into a pattern that shifting the strength of the mean energy 
loss per unit length $dE/dx$ to later times leads to an increased ratio of 
out-of-plane vs. in-plane suppression. We will return to this issue in more 
detail later.

\section{Results}
\label{sec5}

In order to illustrate how the differences between the media computed
with different hydrodynamical models are probed by energy loss, we show 
results for $R_{AA}$ for in-plane and out-of-plane emission for various 
collisions centralities, using both the ASW and AdS energy loss frameworks
and comparing with PHENIX data \cite{PHENIX_RAA_phi}. In all calculations, 
a single free parameter $K$ has been adjusted such that a good description 
of $R_{AA}$ in 0-10\% most central collisions is achieved. We therefore 
refrain from showing any results for central collisions, as they are 
virtually identical for all models \footnote{This is related to the point 
made already in \cite{gamma-h} that $R_{AA}$ is practically insensitive to 
the functional form of the energy loss probability distribution 
$P(\Delta E)$ beyond its first moment.}.

\begin{figure*}[htb]
\epsfig{file=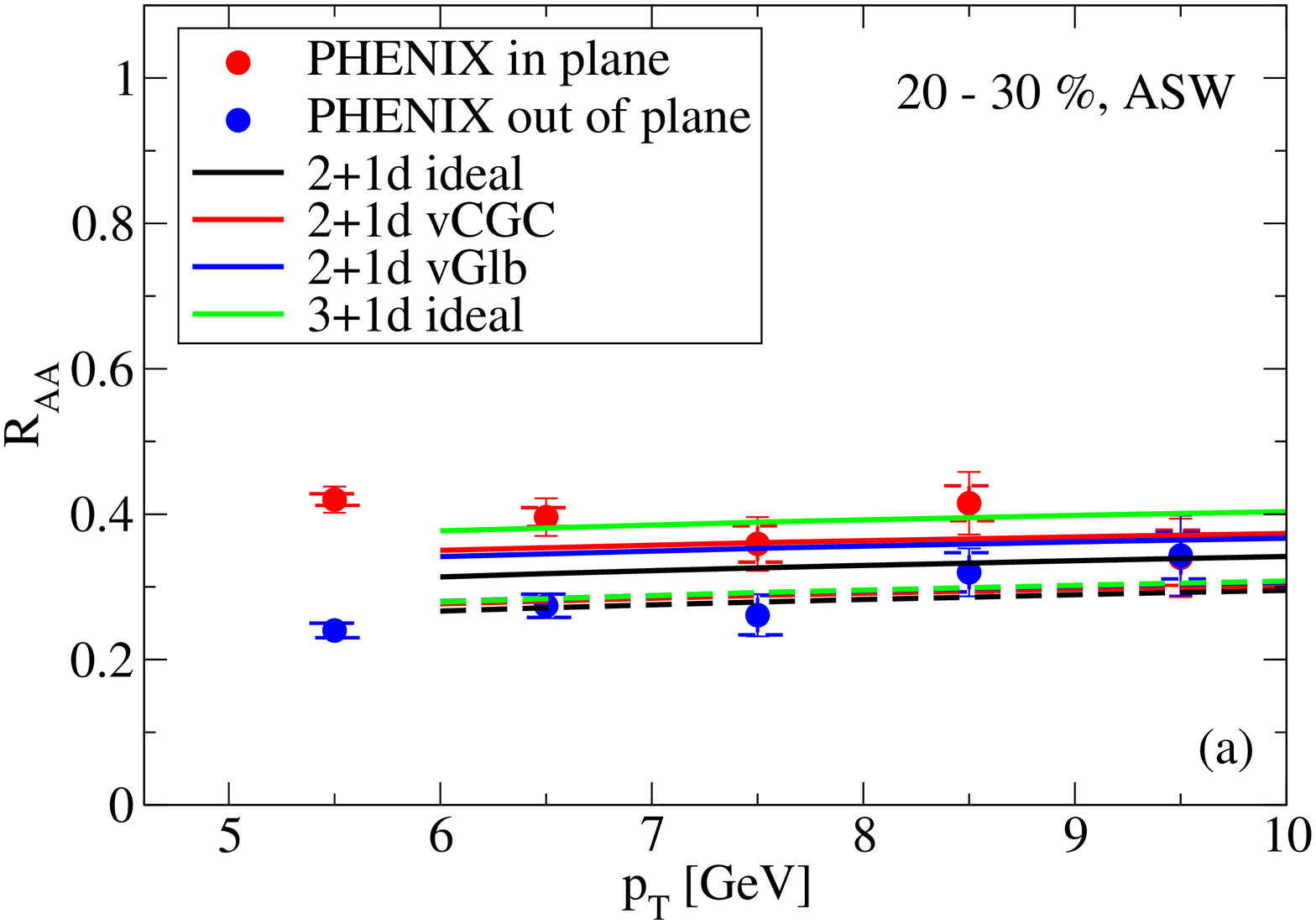, width=8cm}%
\epsfig{file=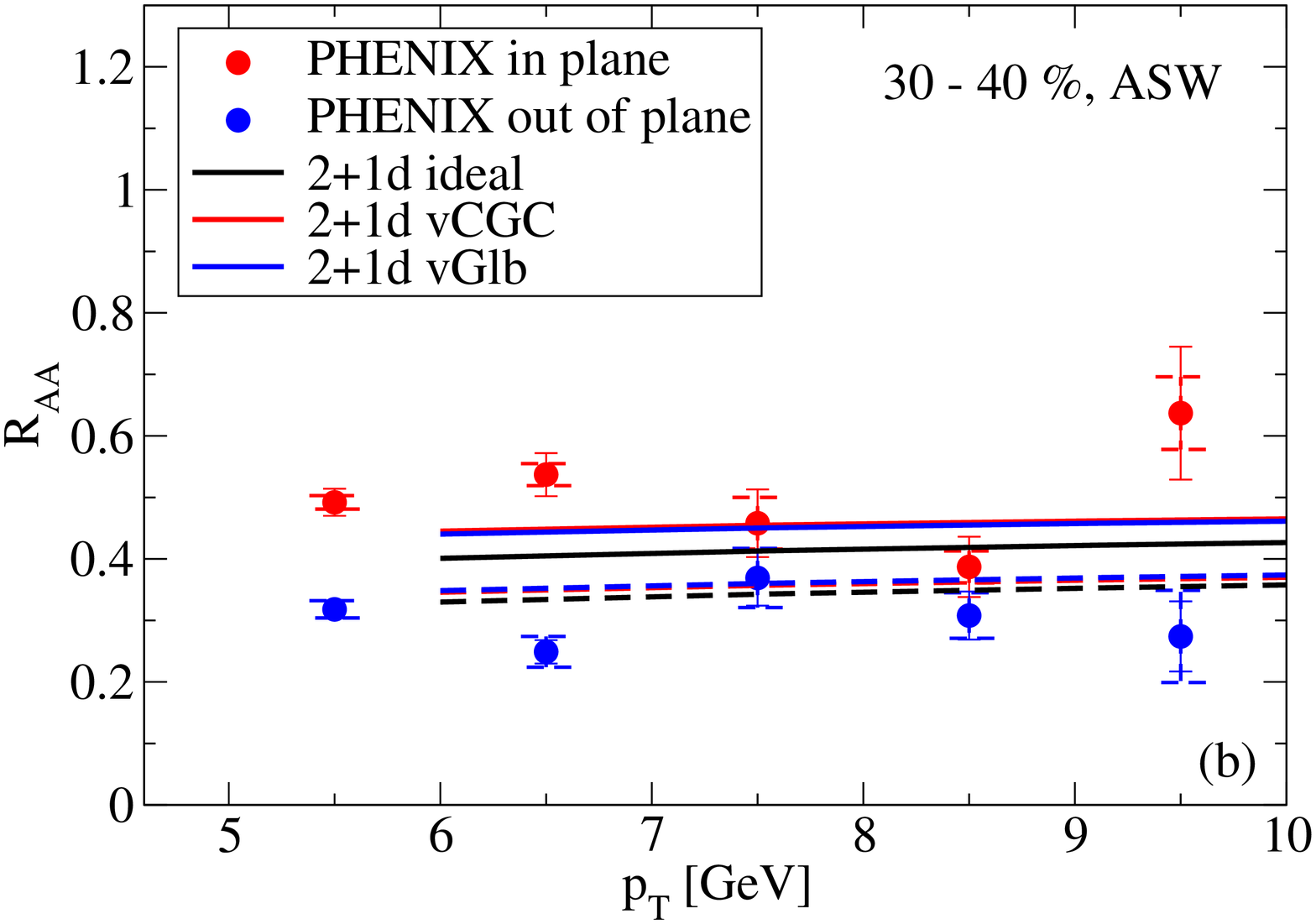, width=8cm}\\
\epsfig{file=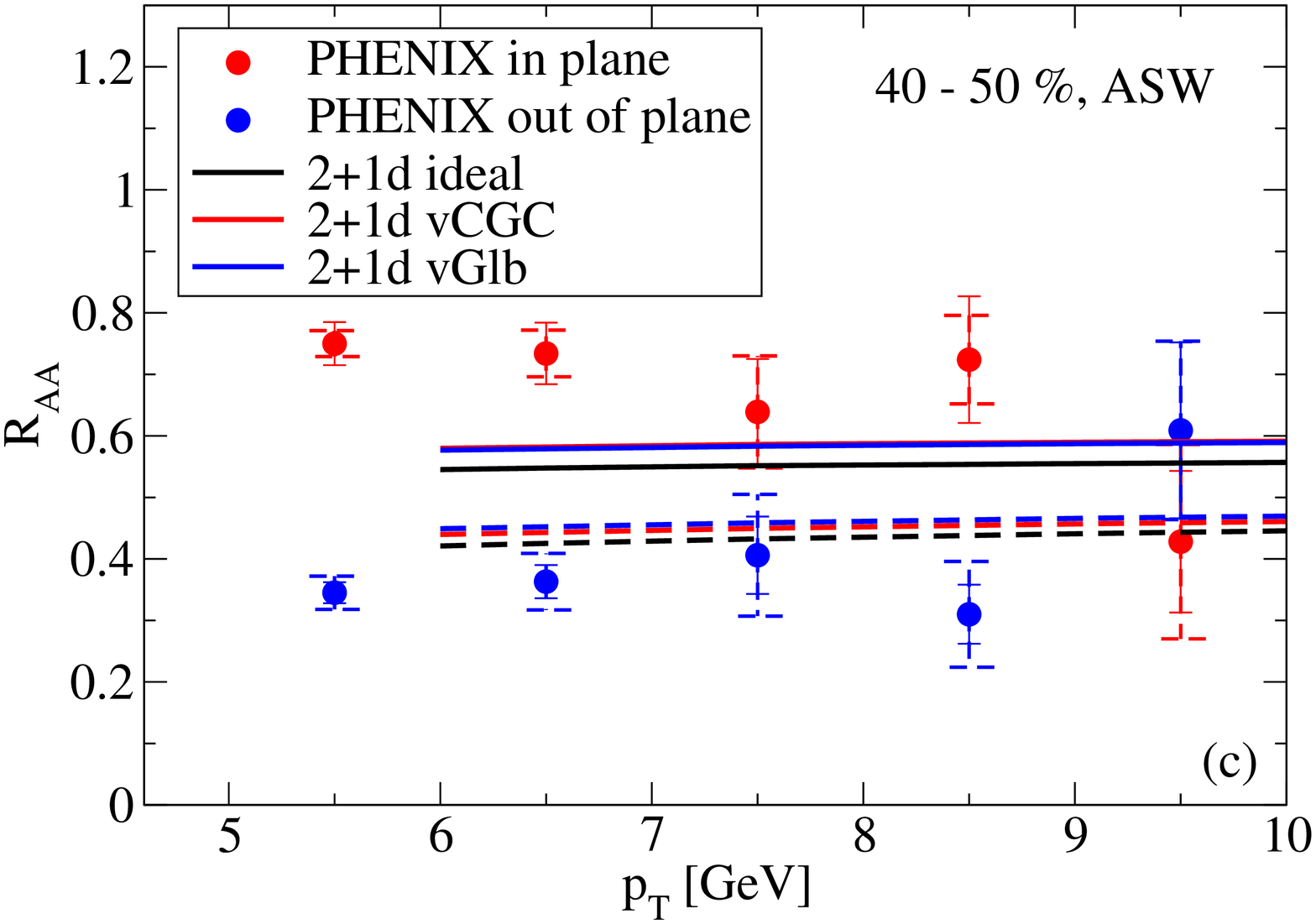, width=8cm}%
\epsfig{file=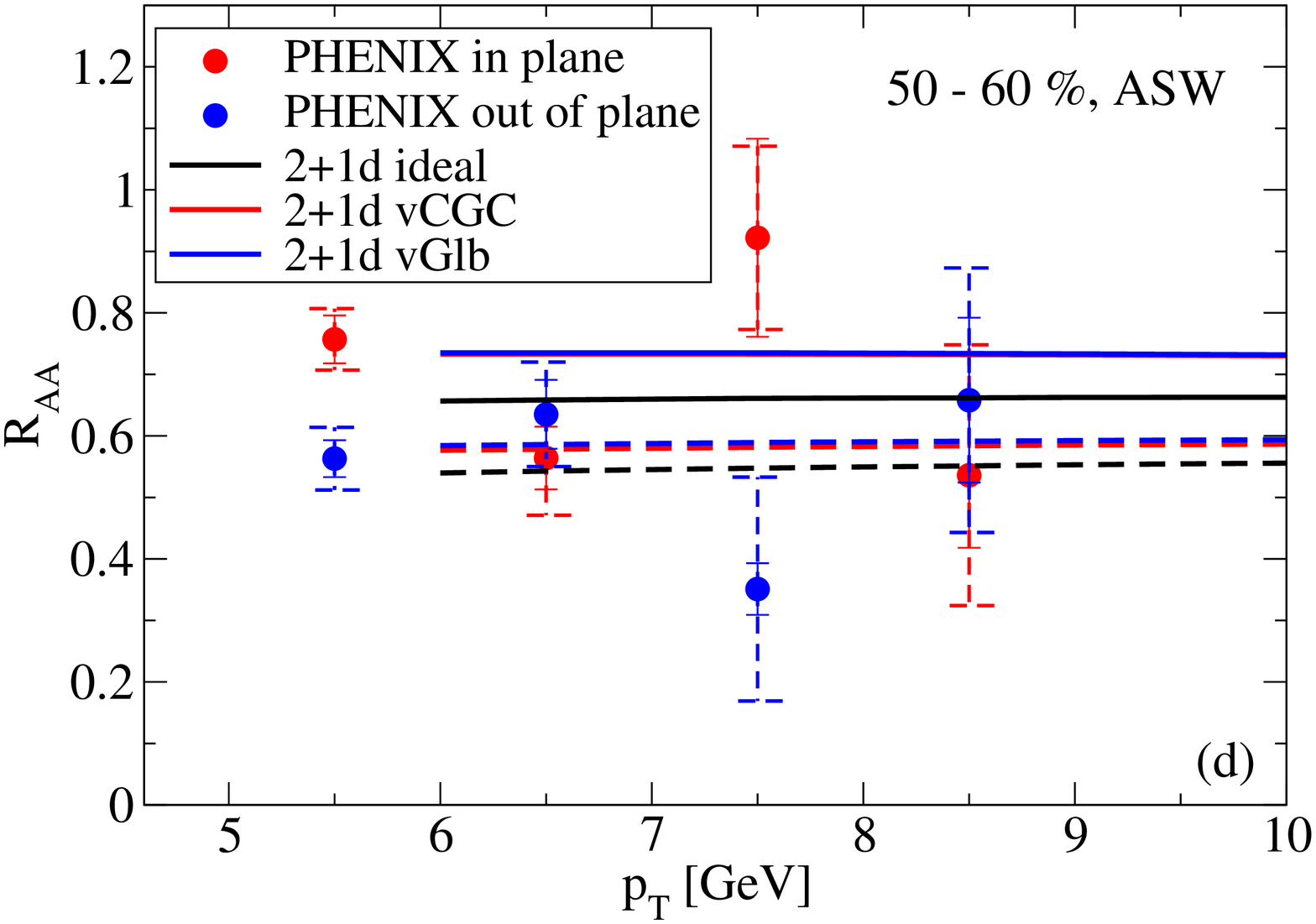, width=8cm}
\caption{\label{F-1} (Color online) The nuclear suppression factor $R_{AA}$ as a function 
of $P_T$, shown in plane (solid) and out-of-plane (dashed). The calculation
was done with the  ASW perturbative radiative energy loss model for 
different hydrodynamical descriptions for the medium, describing $200\,A$\,GeV
Au-Au collisions in four different centrality classes. PHENIX data 
\cite{PHENIX_RAA_phi} are shown for comparison.}
\end{figure*} 

In Fig.~\ref{F-1} we show results for the ASW energy loss models from 
mid-peripheral collisions in the 20-30\%, 30-40\%, 40-50\% and 50-60\% 
centrality classes. While all hydrodynamical models reproduce well
the centrality dependence of the angular averaged (mean) $R_{AA}$ value,
showing very weak $P_T$-dependence, the spread between in-plane and 
out-of-plane emission appears generally too small (with the possible 
exception of the (3+1)-d ideal fluid model). This is especially apparent 
in the 40-50\% centrality class. Note that the successful description of
the centrality dependence of the mean $R_{AA}$ is not trivial, since this 
probes the path length dependence of energy loss: in an elastic energy loss 
model with linear $L$ dependence, even the mean $R_{AA}$ does not correctly
extrapolate from central to peripheral collisions \cite{Elastic2}.

\begin{figure*}[htb]
\epsfig{file=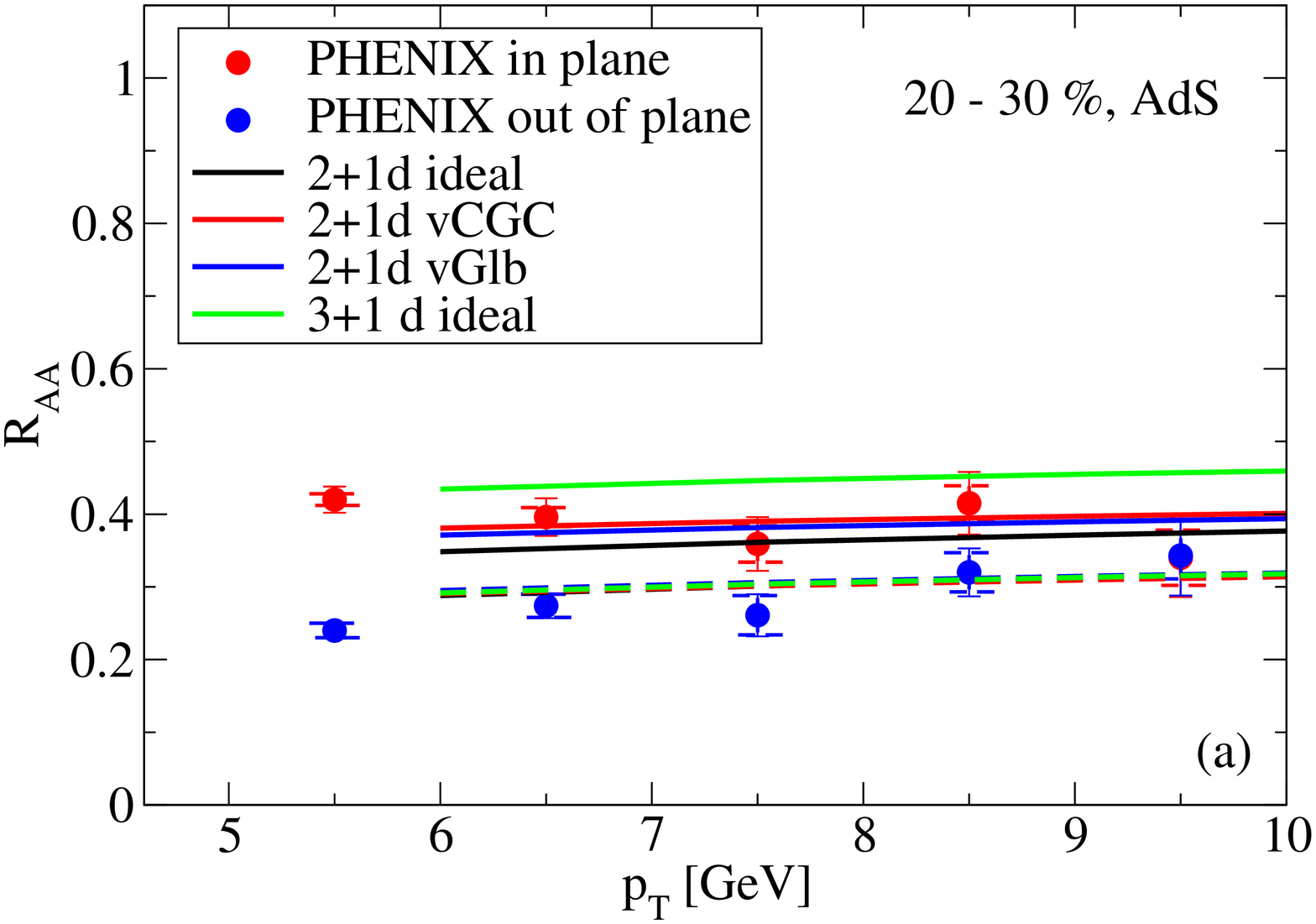,width=8cm}%
\epsfig{file=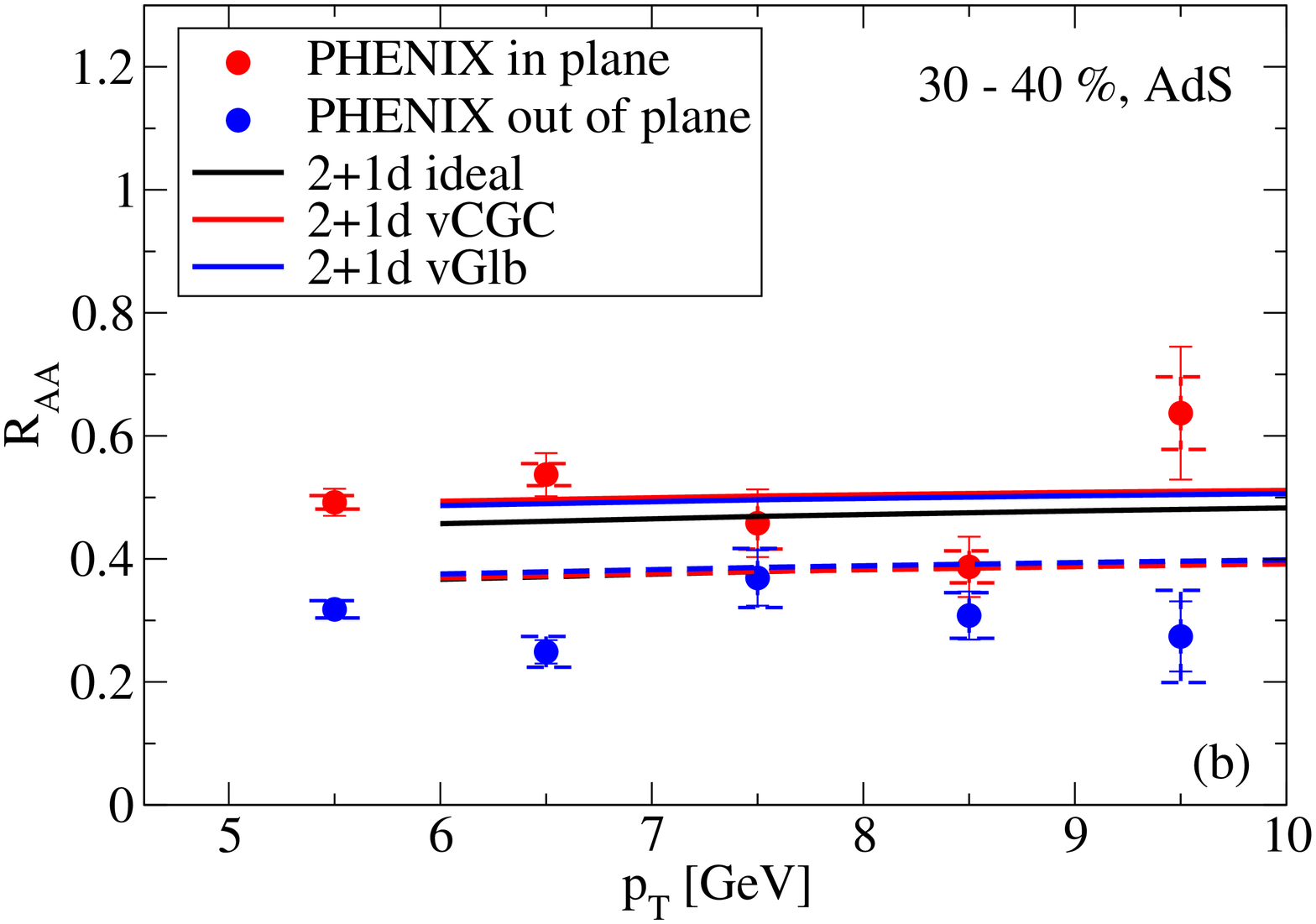,width=8cm}\\
\epsfig{file=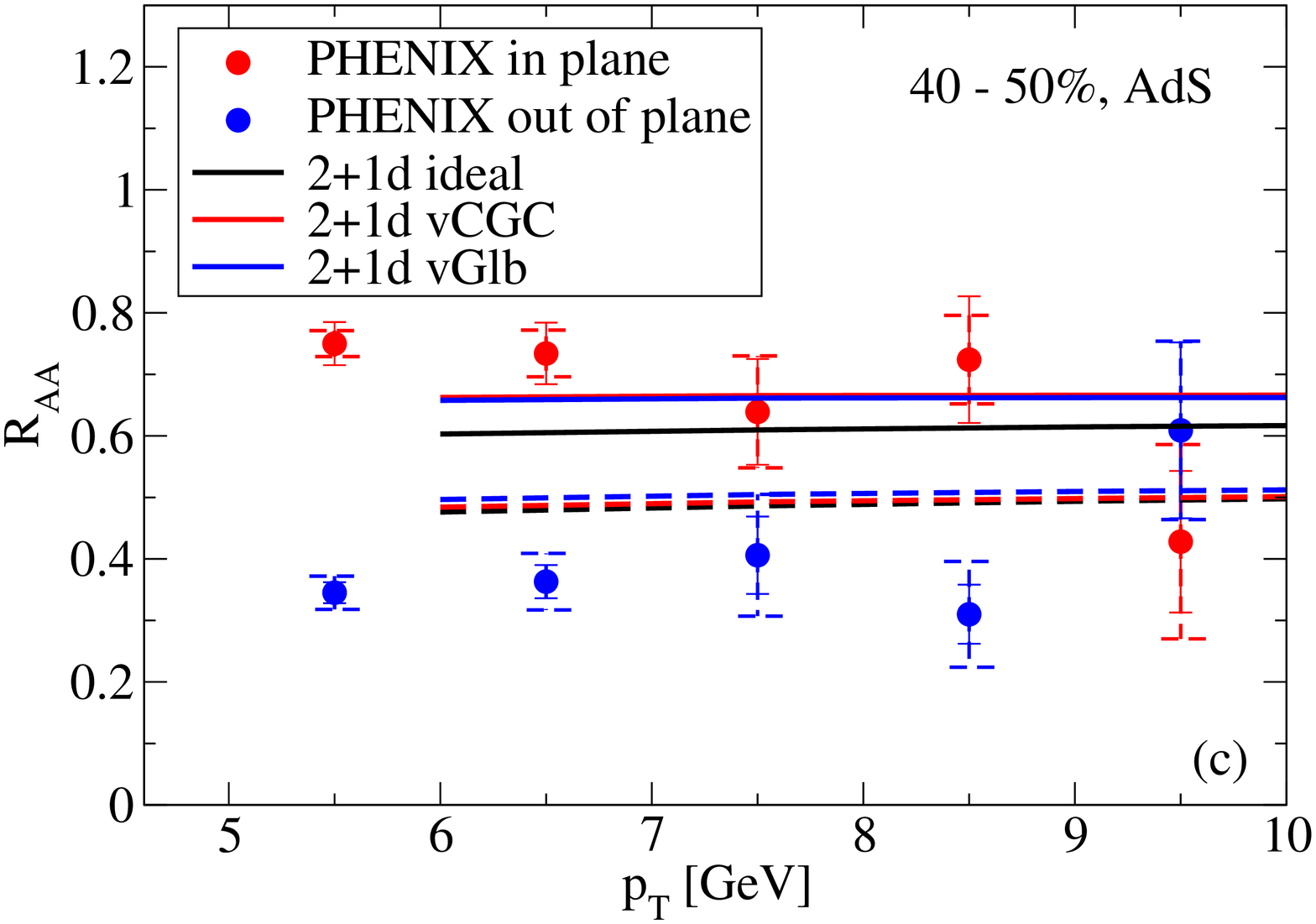,width=8cm}%
\epsfig{file=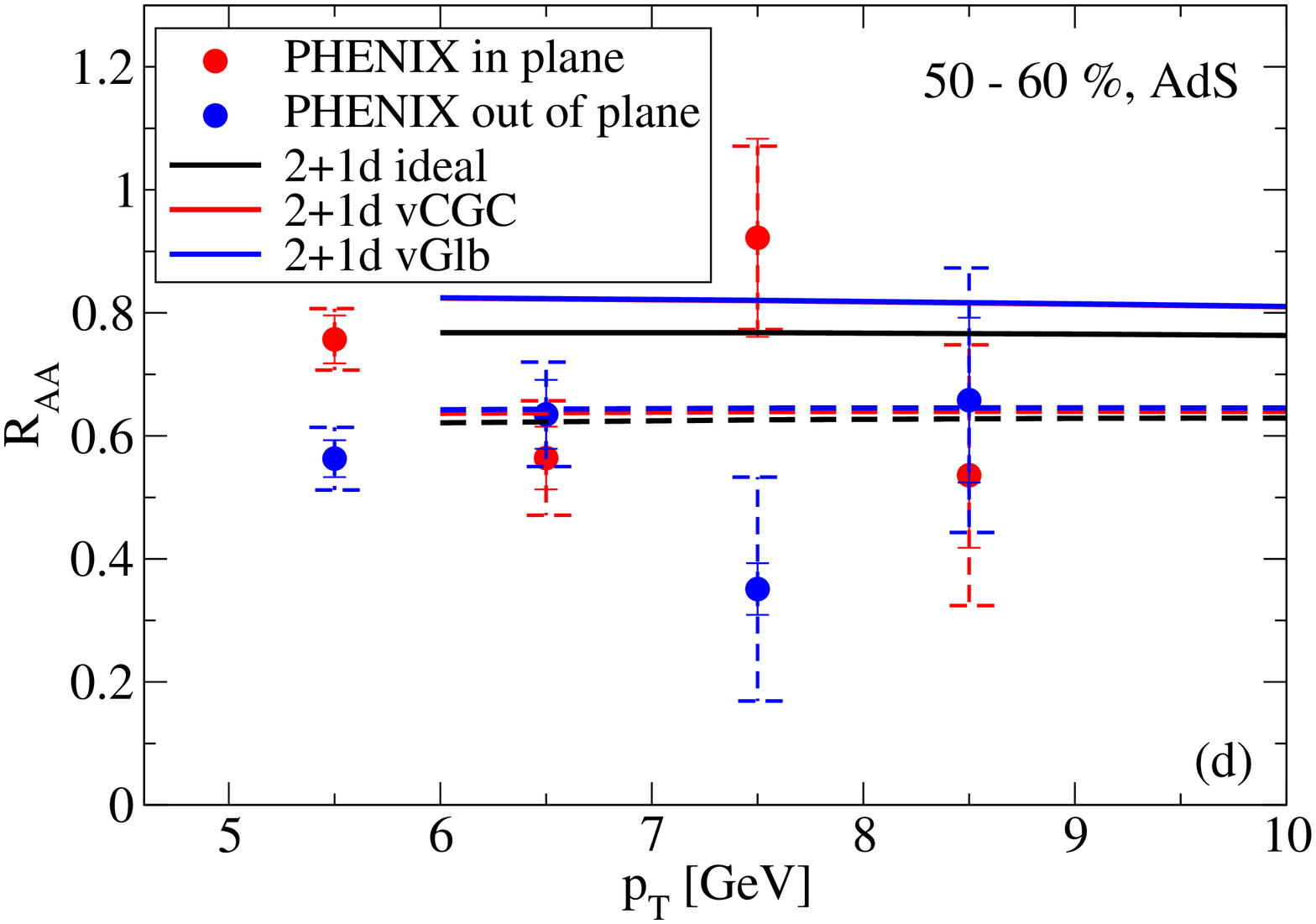,width=8cm}
\caption{\label{F-2} (Color online) Same as Fig.~\ref{F-1} except that the calculation
was done with the AdS strong coupling radiative energy loss model.
}
\end{figure*} 

Fig.~\ref{F-2} shows a similar comparison as Fig.~\ref{F-1} for the strong 
coupling AdS energy loss model which features a stronger $L$ dependence than 
the ASW model ($\sim L^3$ instead of $\sim L^2$). Here there is a tendency 
for the mean $R_{AA}$ to lie above the more peripheral data, but the spread 
between in-plane and out-of-plane emission is generically larger and agrees 
better with the data.

Let us comment on some  trends: In all cases, the in-plane emission shows 
a stronger dependence on the underlying hydrodynamical model than the 
out-of-plane emission. This is to be expected, as the conditions for 
emission out-of-plane (e.g. in terms of average in-medium path length) 
are always more similar to the conditions in central collisions where 
in all models $K$ has been adjusted to describe the data, whereas the 
conditions for emission in the reaction plane change more strongly with 
centrality.

The relatively small difference between fireballs resulting from CGC and 
Glauber initial states is rather remarkable. Given the stronger spatial 
anisotropy of the CGC initial state, one would expect to see this reflected 
in the anisotropy in high $P_T$ parton energy loss. The solution to this 
puzzle seems to lie in the observation made earlier that energy loss 
probes the system at a timescale of $\sim 3-4$ fm, i.e. after the stronger 
pressure gradients of the CGC initial state already had some time to 
reduce the fireball eccentricity. 

With regard to the magnitude of the spread between in-plane and out-of-plane 
emission, we observe that consistently the (3+1)-d ideal hydro leads to the 
largest spread, followed by the viscous hydro models with CGC and Glauber 
initial conditions, while the (2+1)-d ideal fluid medium results in the 
smallest spread. We stress at this point that this observation cannot be 
directly linked to explicit modeling of the dynamics in $z$-direction or 
to viscosity. We will discuss the causes for this ordering in the next 
section.

\begin{figure*}[htb]
\epsfig{file=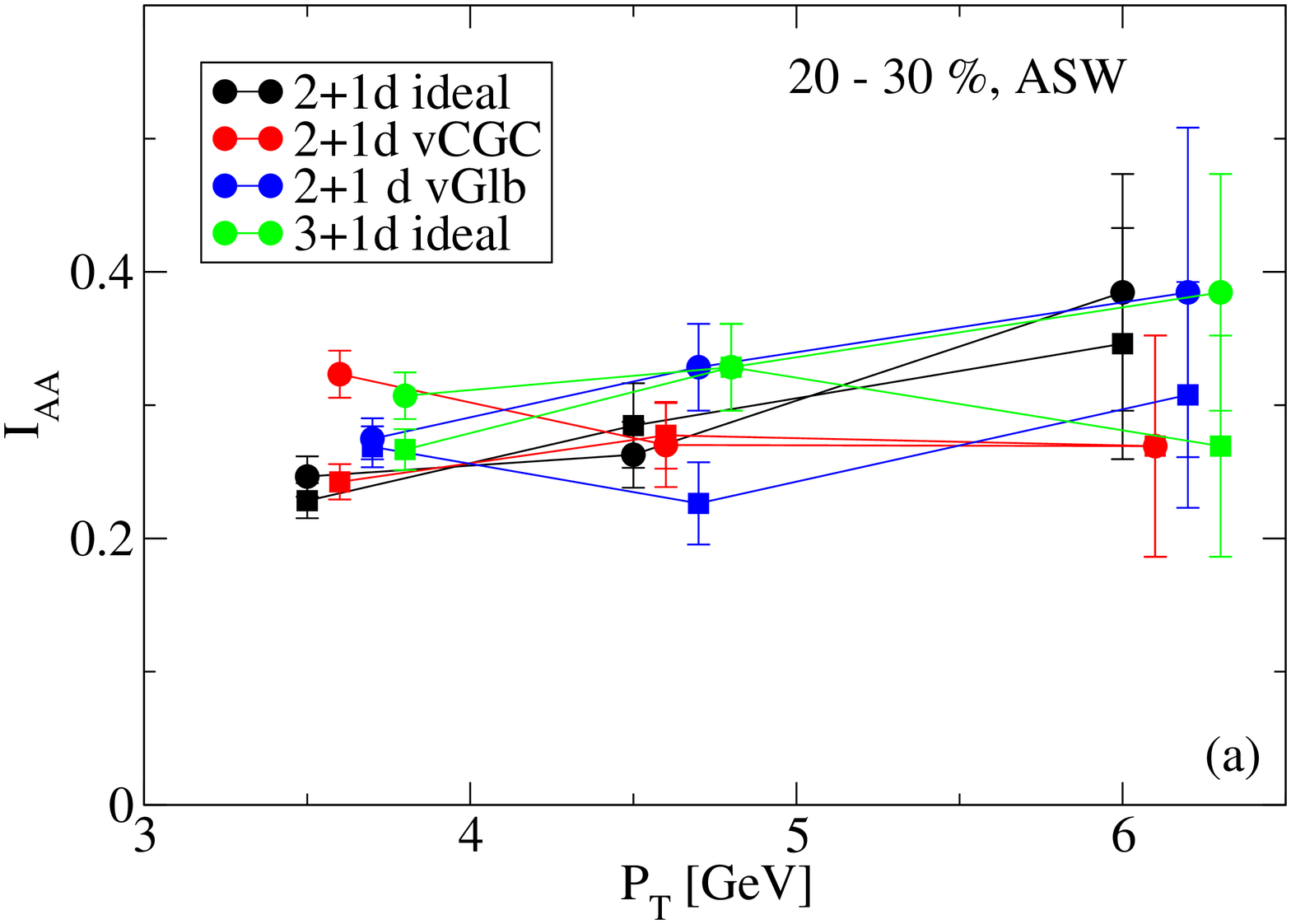,width=8cm}\,
\epsfig{file=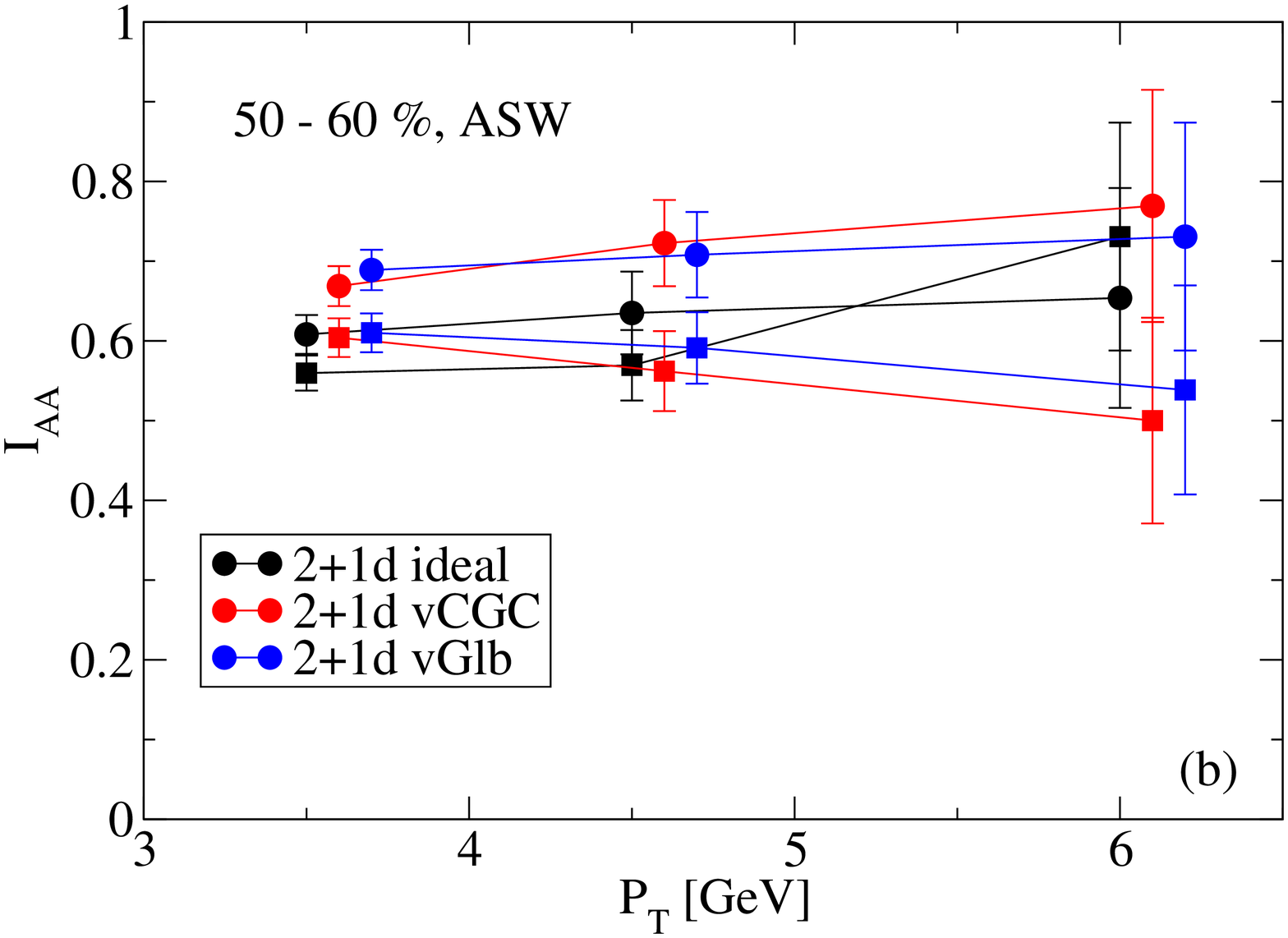,width=8cm}
\caption{\label{F-3} (Color online) The suppression factor $I_{AA}$ of the 
away-side per-trigger yield, calculated in the ASW perturbative radiative 
energy loss model for 4-7 GeV trigger momentum and shown as a function of 
the away-side momentum $P_T$. Squares show $I_{AA}(\phi{\,=\,}0)$ (in-plane), 
circles show $I_{AA}\left(\phi{\,=\,}\frac{\pi}{2}\right)$ (out-of-plane). 
Calculations are done for four hydrodynamical models for $200\,A$\,GeV Au-Au 
collisions for two different centralities.
}
\end{figure*}

In Fig.~\ref{F-3} we show the suppression factor $I_{AA}$ for the away-side 
yield in triggered back-to-back correlations with a trigger momentum range 
of $4-7$\,GeV, using the ASW model. Note that the trigger range is 
\emph{not yet} in a region where hadron production is dominated by the 
fragmentation of hard partons. It has been chosen to match with the range 
of an ongoing experimental analysis, keeping the mentioned caveat in mind.

For 20-30\% centrality, our statistics is not good enough to even cleanly 
separate in-plane from out-of-plane emission. In the 50-60\% centrality 
class, however, we essentially recover the ordering between models in the 
in-plane vs. out-of-plane spread observed before.

\begin{figure*}[htb]
\epsfig{file=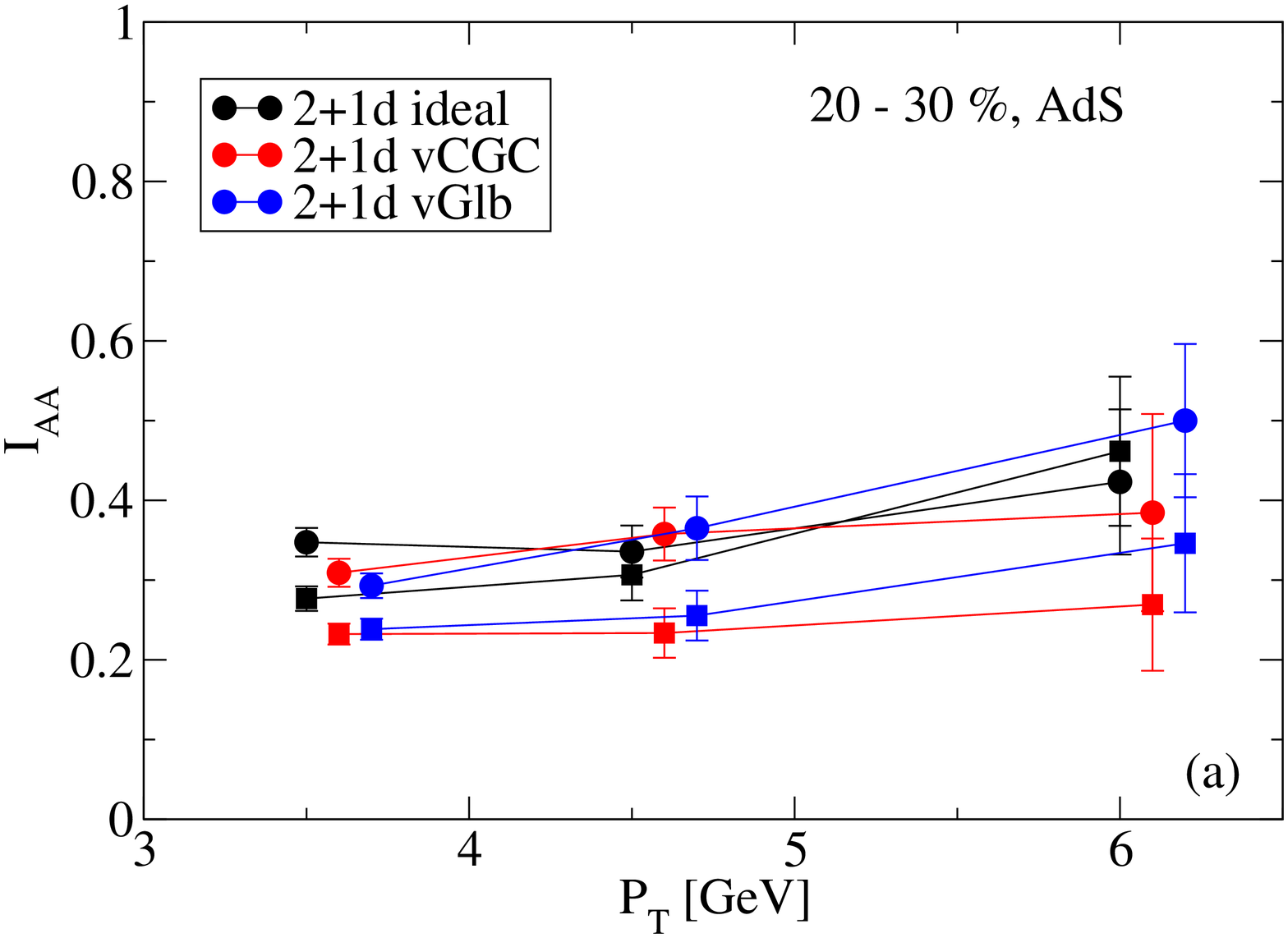,width=8cm}\,
\epsfig{file=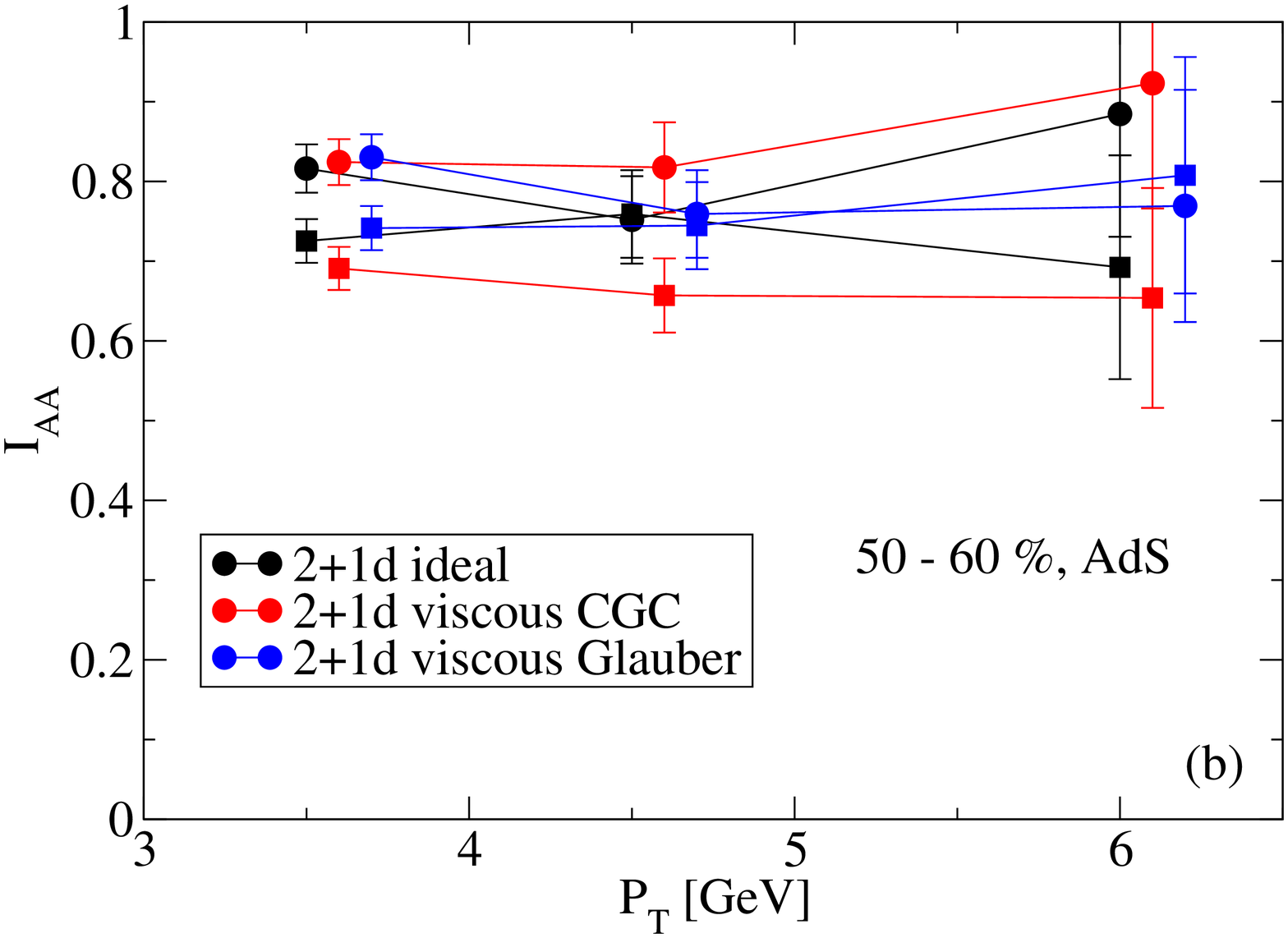, width=8cm}
\caption{\label{F-4} (Color online) Same as Fig.~\ref{F-3} but for the AdS 
strong coupling radiative energy loss model.
}
\end{figure*}

Going to the AdS model in Fig.~\ref{F-4}, the overall magnitude of $I_{AA}$ 
is somewhat different, but qualitatively the picture of the relative 
ordering of the spread between the different models remains unchanged. 
Given that computations of $I_{AA}(\phi)$ are rather involved and suffer 
from limited statistics, it is not clear that $I_{AA}$ offers any real 
tomographic benefit over $R_{AA}$. 

\begin{figure*}[htb]
\epsfig{file=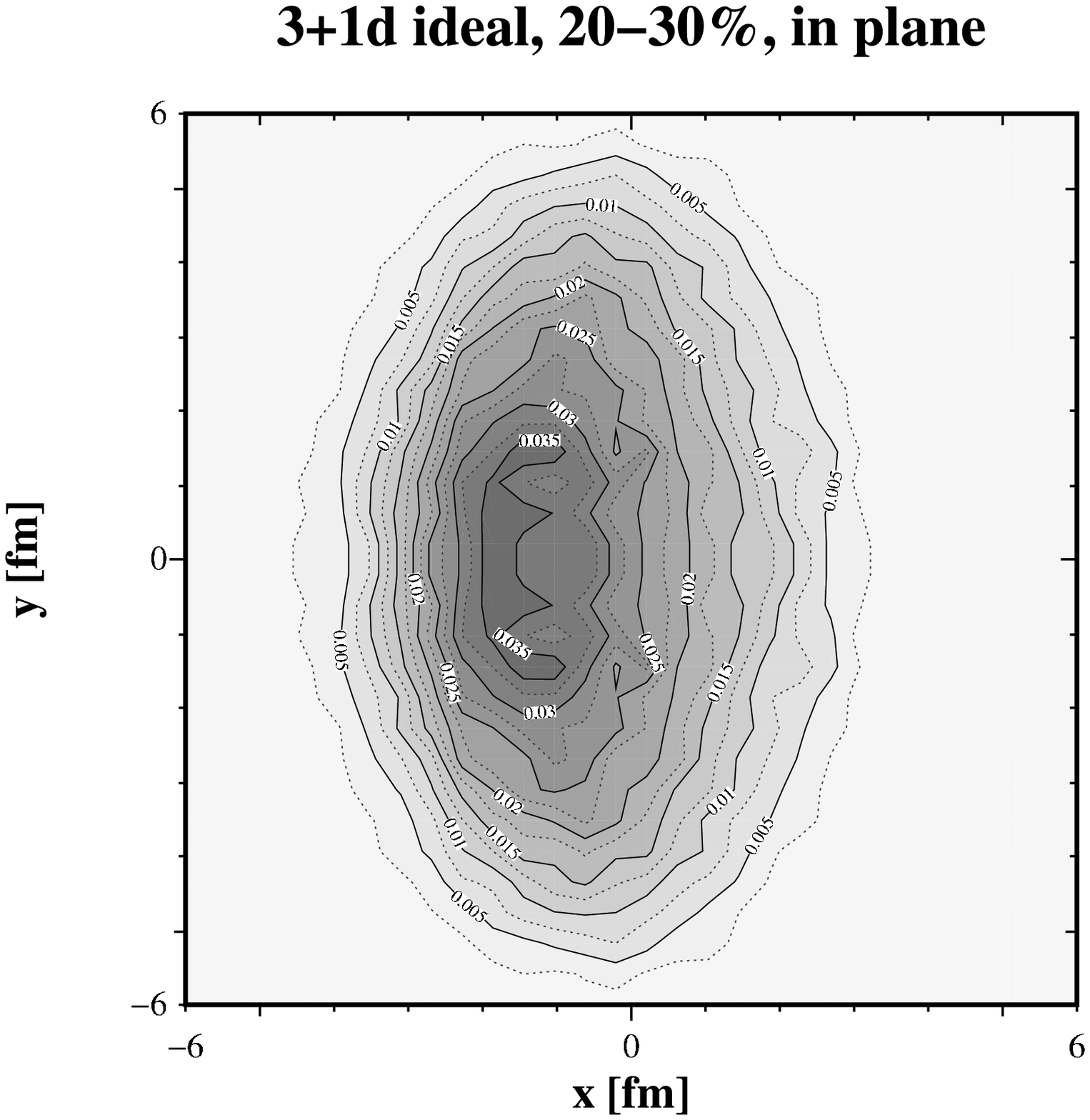,height=8cm} \,\epsfig{file=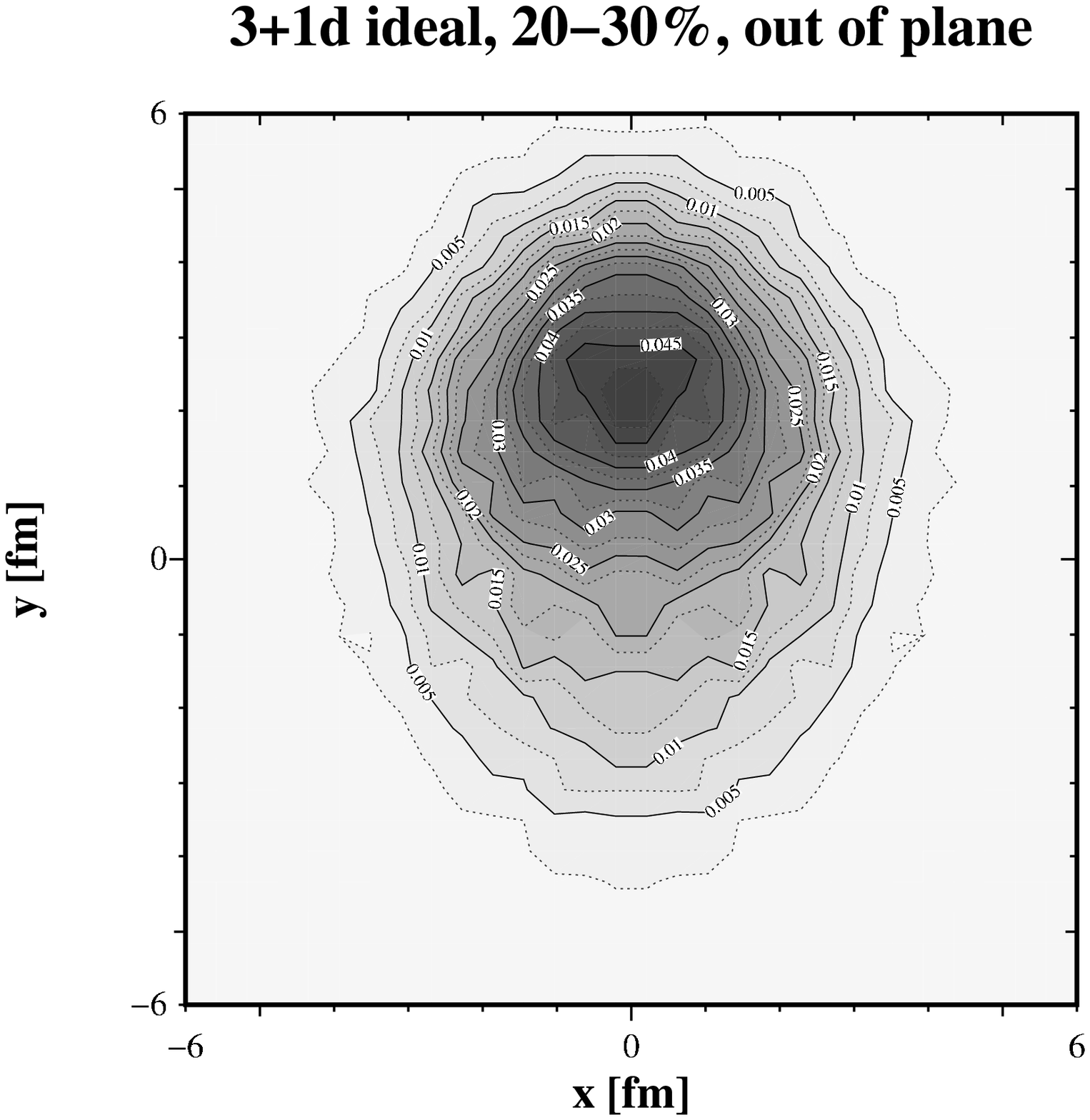,height=8cm}
\epsfig{file=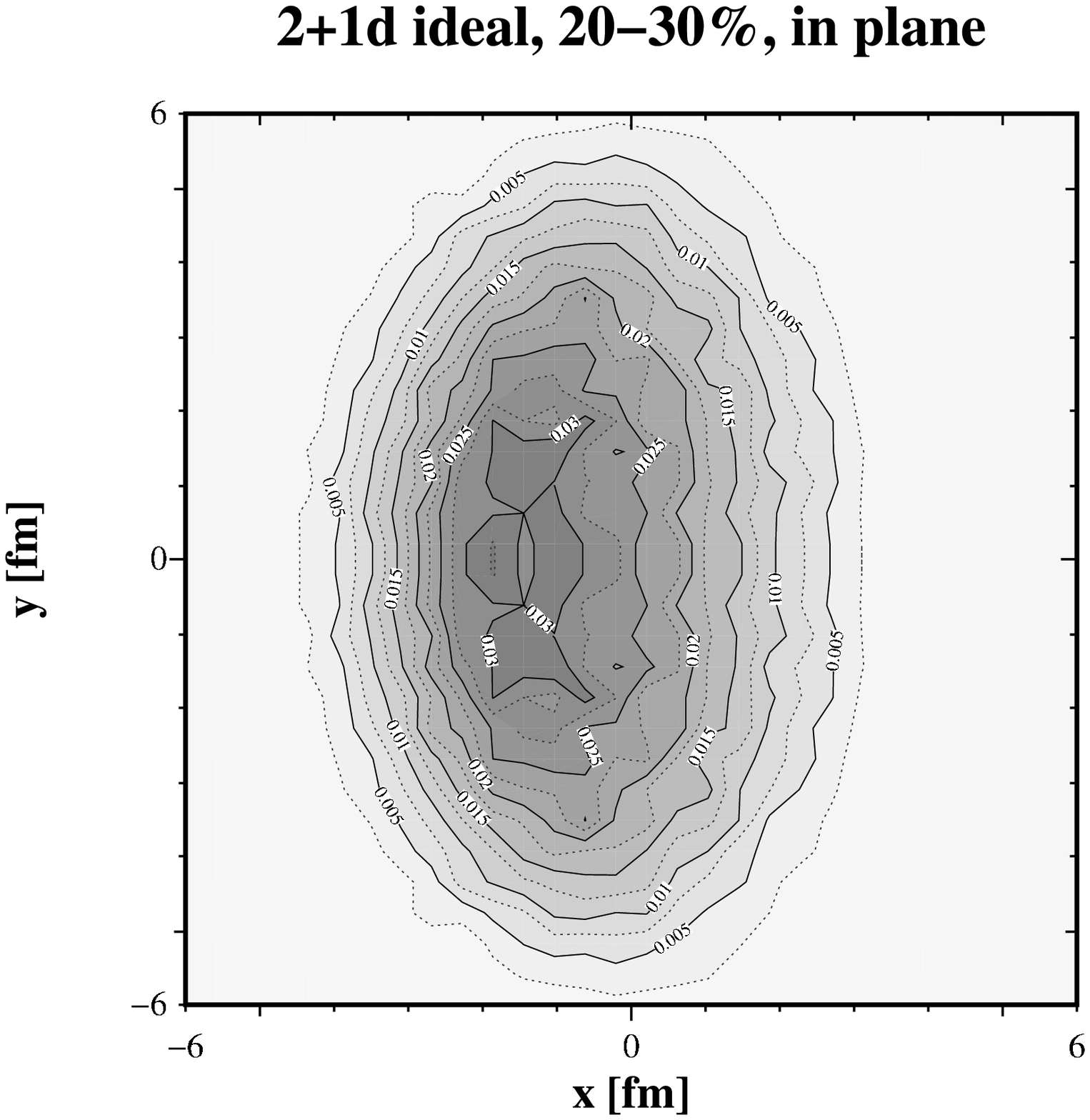,height=8cm} \,\epsfig{file=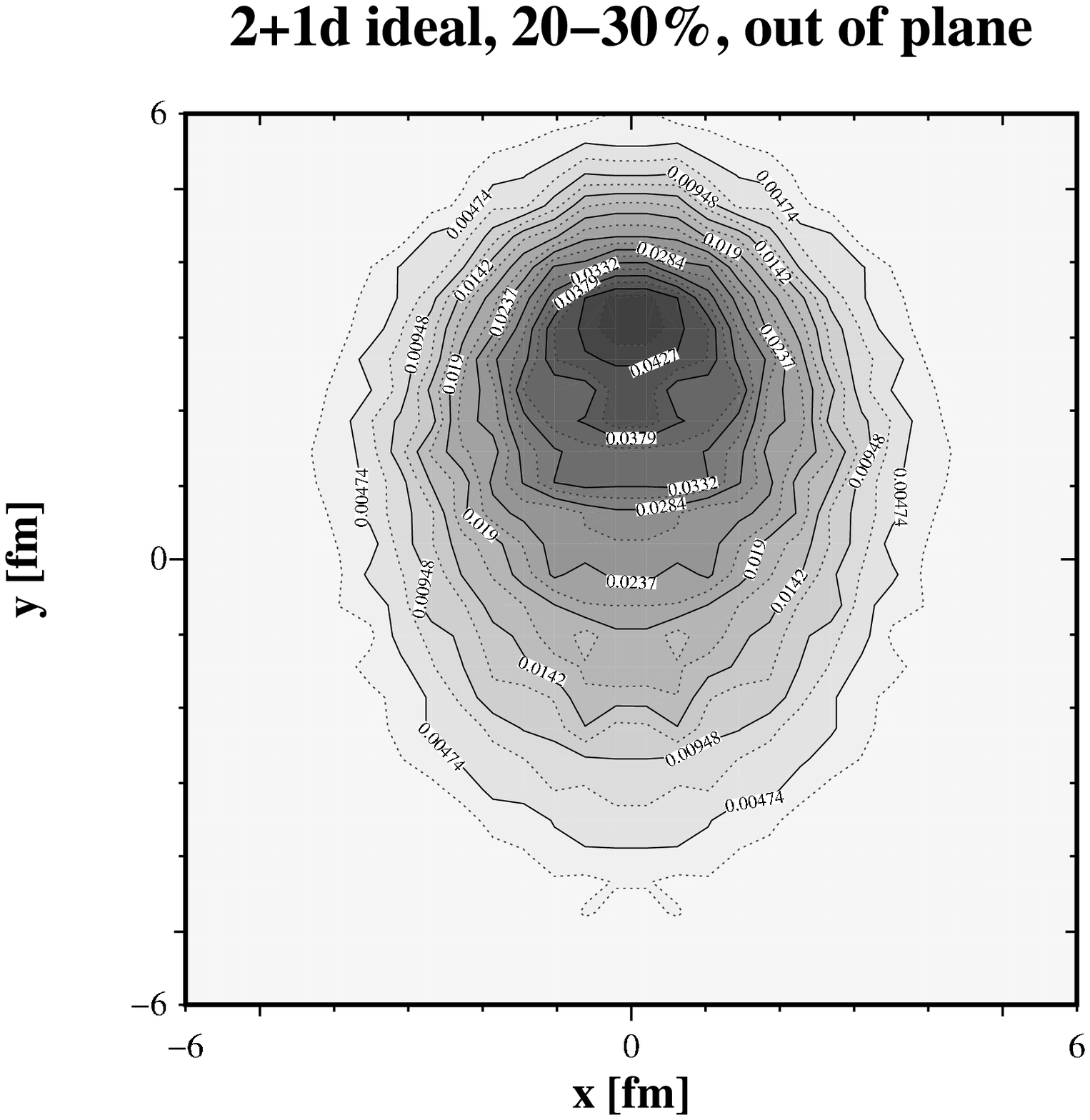,height=8cm}
\caption{\label{F-5}
Probability density of finding a parton production vertex at $(x,y)$ in the 
transverse plane, given an observed hard hadron with 4\,GeV$< P_T <$7\,GeV.
Calculations using the ASW energy loss model are shown for the (3+1)-d (top
row) and (2+1)-d ideal hydrodynamical models (see text). In the case of 
in-plane emission (left panels) the hadron propagates to the $-x$ direction 
and we use $y\leftrightarrow-y$ symmetrization for the plot. In the case of 
out-of-plane emission the hadron propagates to the $+y$ direction and we used
$x\leftrightarrow-x$ symmetrization. Countours are at linear intervals.
}
\end{figure*}

\begin{figure*}[htb]
\epsfig{file=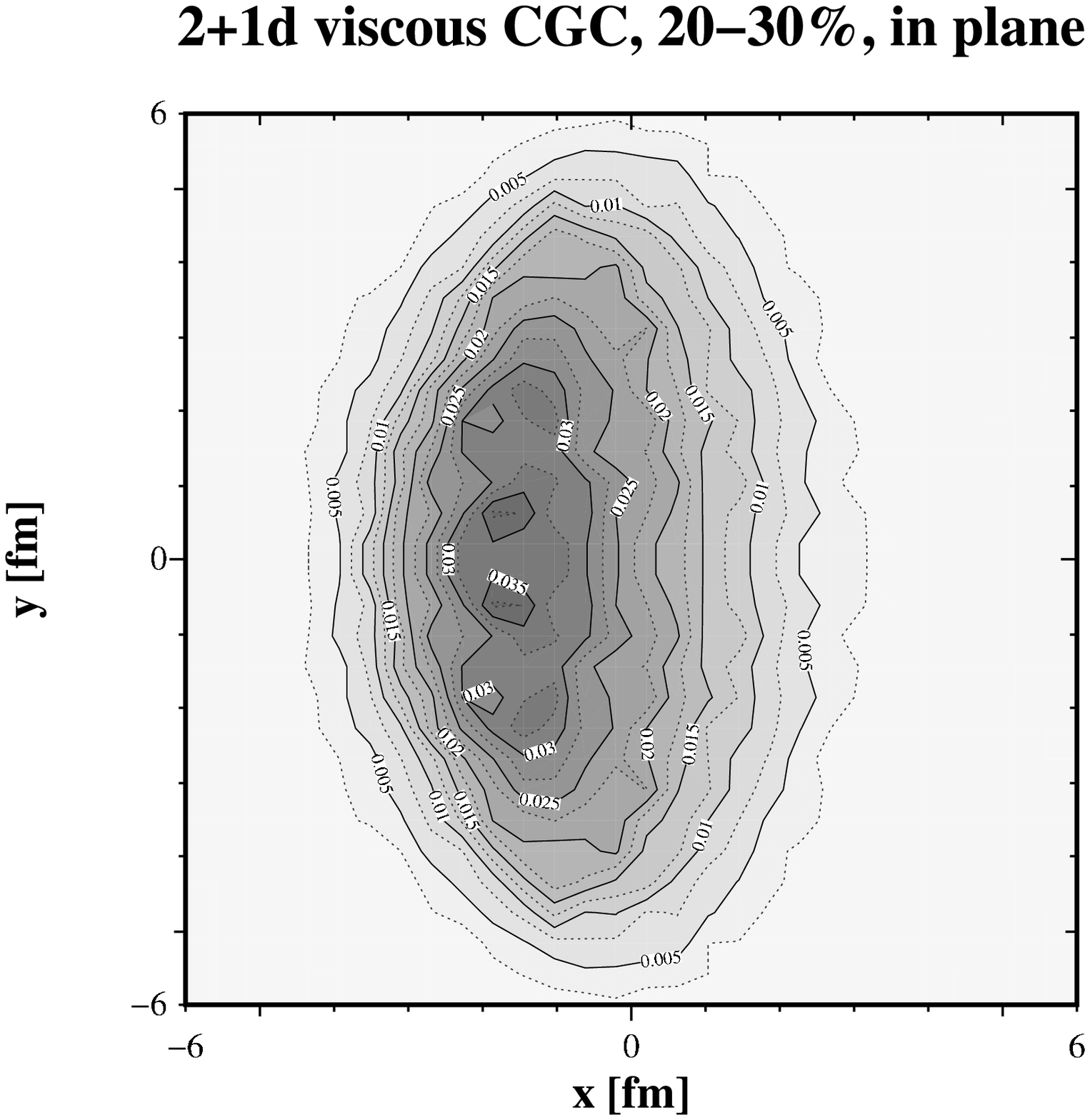,height=8cm} \,\epsfig{file=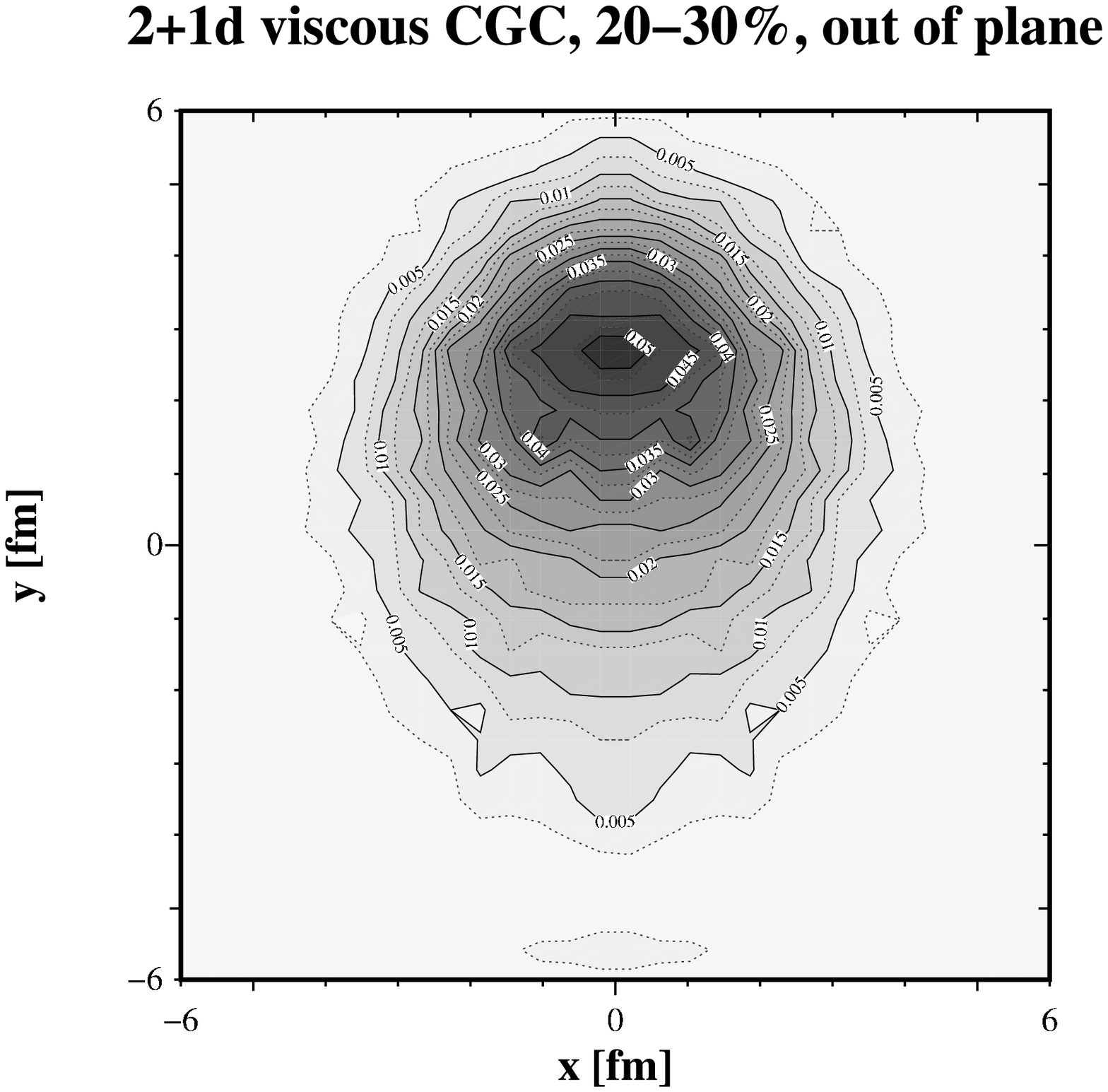,height=8cm}
\epsfig{file=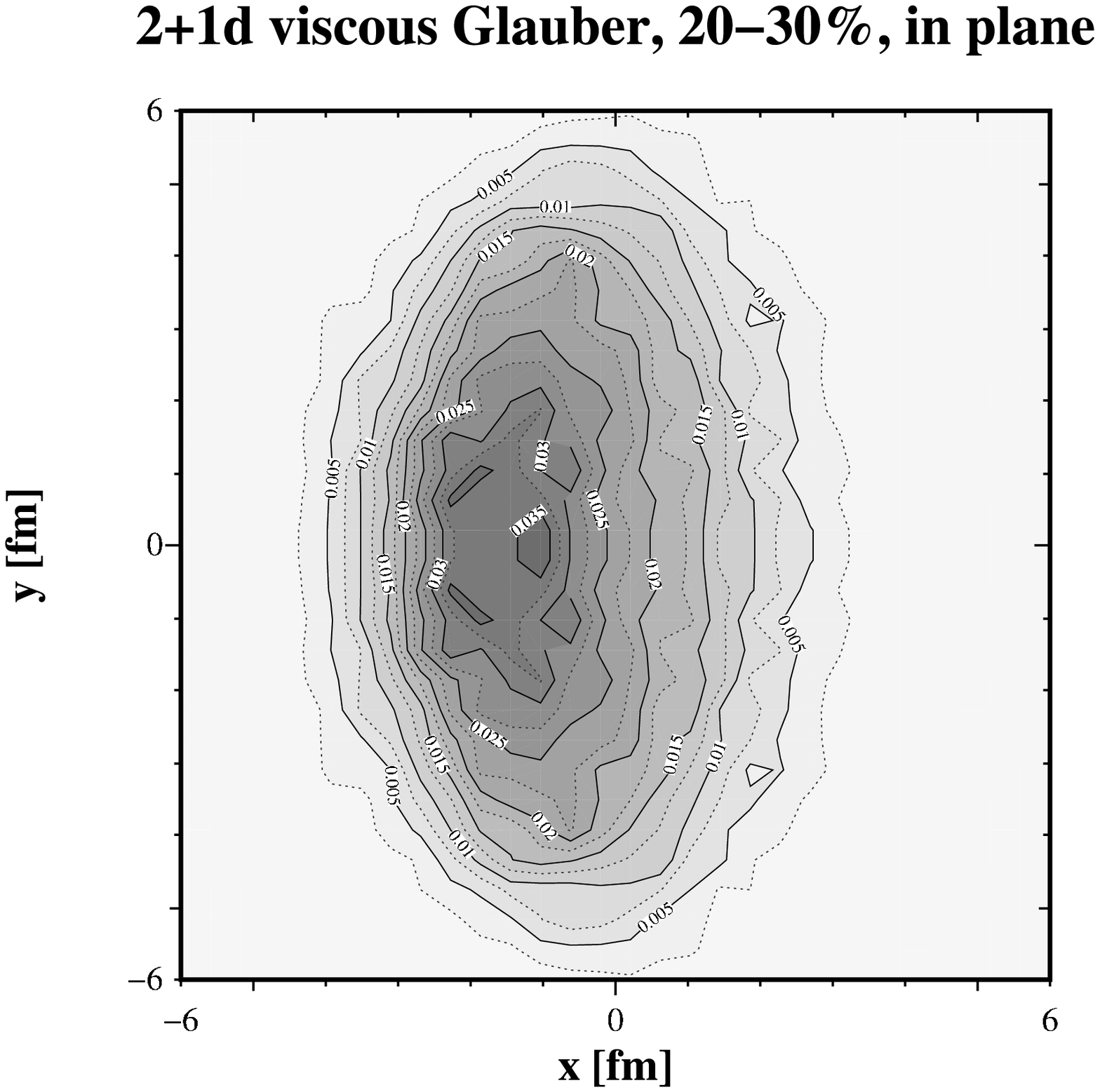,height=8cm} \,\epsfig{file=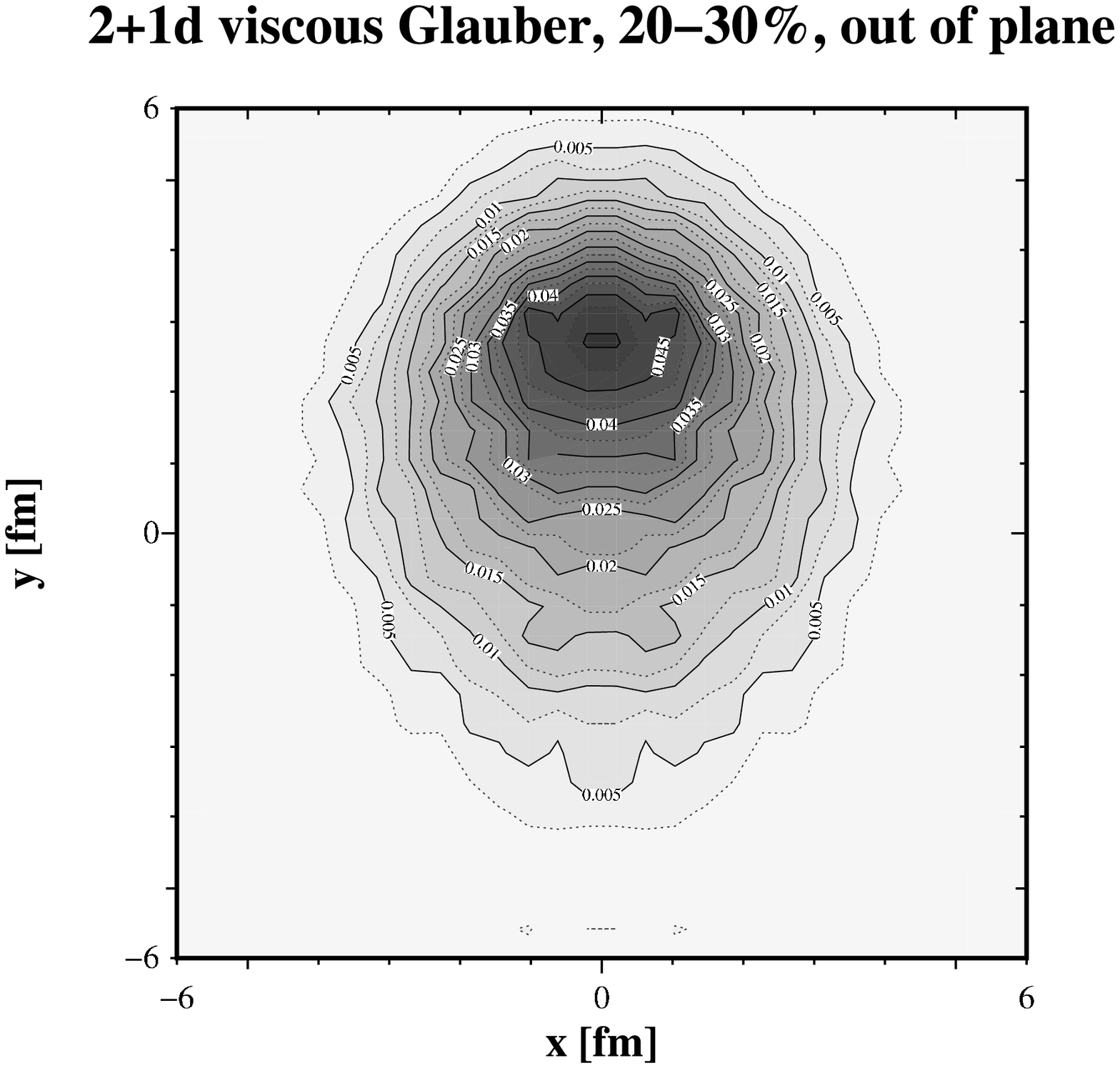,height=8cm}
\caption{\label{F-6} Same as Fig.~\ref{F-5} but for the viscous hydrodynamical
models with CGC-fKLN (top row) and Glauber (bottom row) initial conditions 
(see text).
}
\end{figure*}

In order to illustrate more clearly how the energy loss probes the medium 
density evolution, we show in Figs.~\ref{F-5} and \ref{F-6} the conditional 
probability density to have the production vertex of the hard parton at 
position $(x,y)$, given that a hard hadron was observed. We show calculations
for the 20-30\% centrality class using the ASW energy loss model; the 
qualitative features and differences between different hydrodynamical models 
seen in these plots repeat themselves in other centrality classes and are
slightly enhanced for the AdS energy loss model which we do not show here.

In the absence of a medium, this production vertex density distribution 
is given by the binary collision distribution Eq.~(\ref{E-Profile}). The 
medium biases the distribution in a characteristic way towards the surface, 
as partons produced in the dense medium core are unlikely to escape with a 
significant fraction of their momentum left. The plots have been obtained 
by binning the distribution of triggered events in the MC code for the 
computation of back-to-back hadron correlations, but the same conditional 
$P(x,y)$ for the trigger distribution is also underlying the single hadron 
suppression $R_{AA}$.

As expected, the degree of surface bias is much larger for out-of-plane 
than for in-plane emission, reflecting the larger degree of suppression 
seen out-of-plane. Comparing the $y$-position of the maximal out-of-plane 
emissivity, a stronger degree of surface bias in the (2+1)-d ideal vs. the 
(3+1)-d ideal hydrodynamics is readily apparent. The two viscous models 
reflect their different density evolutions in a less straightforward way, 
for example in the different shape of the emissivity maximum or in the 
distortion of the outer contour lines. As previously observed in 
\cite{Correlations}, there is no evidence for strictly surface-biased 
emission or a corresponding tangential bias. In all models, a significant
fraction of observed hadrons originates from the medium core.

\section{Interpretation of the results}
\label{sec6}

Arguably the most important question in light of the magnitude of the 
measured spread between in-plane and out-of-plane emission is whether
the data require a strong coupling description of energy loss along the 
lines of the AdS model or if, given a suitable description of the medium, 
perturbative QCD is able to account for the data. A related issue of
similar importance is what constraints for hydrodynamical models can be 
derived from measurements of hard probes that cannot also be gained from 
bulk matter data.

Let us discuss these questions in view of our findings. It is a fortunate 
accident that the Cooper-Frye surfaces (see Fig.~\ref{surfaces}) of all 
(2+1)-d models studied here are almost identical. Since we can easily test 
how $R_{AA}(\phi)$ is changed when we make $\tau_0$ (here denoting the
starting time for energy loss) equal in these runs (cf. 
Fig.~\ref{F-line-integrals}), we have practically eliminated any 
effect of different in-medium path lengths and can ascribe any differences 
in the resulting suppression patterns to the density distributions probed 
along the hard parton path. 

We note that the (2+1)-d ideal and vGlb hydrodynamics both share (to
good approximation) the same shape of the initial state whereas the 
vCGC hydrodynamics has a different initial state. This enables us to 
disentangle different contributions to the in-plane vs. out-of-plane 
spread: about 50\% of the difference in spread between the (2+1)-d ideal 
and vCGC hydrodynamics shown in Fig.~\ref{F-1} can be ascribed to the 
difference in the initial time $\tau_0$ (Fig.~\ref{F-line-integrals} 
indicates that a small $\tau_0$ is strongly disfavoured), about 35\% of 
the difference results from viscosity, and the remaining 15\% are caused 
by the difference between CGC and Glauber initial profiles.

Turning to the (much larger) spread observed in the (3+1)-d hydrodynamics, 
we can surmise that most of the increase must arise from the different size 
of the freeze-out hypersurface (and the resulting different distribution of 
in-medium path lengths). This in turn implies that, once the parameter $K$ 
has been adjusted to yield the same mean $R_{AA}$ as in the other models,
the numerical value of $\hat{q}$ at each space-time point is much smaller 
in the (3+1)-d ideal hydro model than in the other models (see 
Fig.~\ref{F-viscous-entropy}). Another way to state this is that in the 
(3+1)-d model partons travel on average a larger distance before they
acquire the same amount of virtuality transfer from the medium.

The common theme in all these findings is that the spread between in-plane 
and out-of-plane emission is increased whenever energy loss is shifted to 
later times. This may occur due to viscous heating during expansion, or 
in response to the choice of a large $\tau_0$, or as the result of an $L^3$ 
path length weighting as in the AdS model --- while the details differ, the 
net result is qualitatively the same in each of these cases. We note that
this agrees qualitatively with the conclusions in \cite{Liao:2008dk} whose
authors achieve a large in-plane vs. out-of-plane spread by requiring the
hard parton to suffer the largest energy loss rate in a relatively thin 
shell of matter whose temperature is close to $T_c$.

This observed connection between the time dependence of energy loss and the 
in-plane vs. out-of-plane spread requires an explanation, especially in 
light of the fact that elliptic flow {\em decreases} the initial spatial 
anisotropy of the system over time. The key to understanding this 
phenomenon lies in the realization that the spatial anisotropy exists on 
a scale of the order of the spatial size of the transverse overlap region, 
i.e. several fm. It is thus a global property of the medium. If energy 
loss were significant only at very early times $\tau \alt 1$\,fm/$c$ (as 
approximately true for elastic energy loss), it could not resolve any 
phenomenon on a distance scale $d \gg 1$ fm and would thus probe the 
medium properties only locally. In other words, in such a model most high 
$p_T$ partons would be blind to the spatial anisotropy (see discussion 
in \cite{Elastic1}), and consequently the observed spread would be very 
small.

While the above argument is strictly true only for a homogeneous medium, 
it still holds qualitatively for an inhomogeneous and hydrodynamically 
evolving medium: The spread between in-plane and out-of-plane emission 
increases not because energy loss is shifted to late times, but because 
the typical time scale over which energy loss is strong approaches the 
global spatial scales of the medium, and hence the partons are increasingly 
able to probe the medium globally.

Unfortunately this finding does not translate into straightforward 
constraints for the medium evolution model. Rather than being dominated 
by a single large effect, $R_{AA}(\phi)$ appears to be sensitive to a 
number of effects of roughly equal importance. However, while some 
constraints are qualitatively similar to what was found in the soft 
sector (for example, a CGC initial profile causes both a larger $v_2$ 
in the soft sector and a larger azimuthal variation of $R_{AA}$ in the 
hard sector), others are qualitatively different and potentially more 
valuable (viscosity causes smaller $v_2$ for the bulk matter but a 
larger azimuthal variation of $R_{AA}(\phi)$).

Given that our results do not exhaust the parameter space of hydrodynamical 
evolutions compatible with bulk data, it is entirely conceivable that 
an evolution can be found for which both magnitude and spread of 
$R_{AA}(\phi)$ are described within pQCD dynamics. Presumably, a viscous 
(3+1)-d hydrodynamics with a CGC initial state, late thermalization and 
low freeze-out temperature would be a good candidate. At this point, we 
see no reason to conclude that the data require a strong-coupling model.

\section{Summary}
\label{sec7}

The goal of this paper was to develop a better understanding of the 
tomographic power of parton energy loss measurements in relativistic
heavy-ion collisions as probes of the medium created in these collisions
and/or of the mechanism by which hard partons lose energy in such a medium.
To this end we presented a study of the dependence on the angle with 
the reaction plane $\phi$ of the nuclear suppression ratio $R_{AA}(\phi)$ 
and the away-side per-trigger yield $I_{AA}(\phi)$ in triggered 
back-to-back correlations, for four different hydrodynamic models of 
the fireball evolution and two different models for the path length 
dependence of parton energy loss. The models were tightly constrained 
by ensuring that parameters were chosen such that the soft-hadron 
transverse momentum spectra are well described for all collision 
centralities, and that all models yield the same nuclear suppression 
factor $R_{AA}$ for pions in central Au-Au collisions. We then studied 
differences in the dependence of $R_{AA}$ and $I_{AA}$ on collision 
centrality, transverse momentum $P_T$, and azimuthal emission angle 
$\phi$ relative to the reaction plane.

We found that, after tuning the models to reproduce the correct pion $R_{AA}$
in central collisions, they all gave approximately identical results
for the collision centrality dependence of the azimuthally averaged
$R_{AA}(P_T)$, featuring weak $P_T$-dependence and yielding good 
qualitative agreement with experimental data. The tomographic power
of this azimuthally averaged quantity for distinguishing between different 
(realistic) medium models is therefore low. A stronger path length 
dependence of the energy loss ($\sim L^3$ instead of $\sim L^2$) 
produces a slightly stronger impact parameter dependence of the 
$\phi$-averaged $R_{AA}$, but small differences between the different
hydodynamic models for the medium interfere with this tendency, making it 
difficult to disentangle these effects. The centrality dependence of
the azimuthally averaged $I_{AA}$ is {\itshape a priori} a bit more promising since it was found 
to react more strongly to changes in the path length dependence of the 
energy loss. On the other hand, to measure this quantity with good
statistical precision is much more difficult. Furthermore, it is not 
obvious from our studies that such a measurement will contribute any
useful information about the medium that would help distinguish between
different hydrodynamic evolution models. The azimuthally averaged $R_{AA}$
is found to provide practically no such discriminating power.

The measurement of the in-plane vs. out-of-plane variation of $R_{AA}(\phi)$
provides much better discrimination \footnote{Whether the same holds true 
 for the azimuthal variation of $I_{AA}$ depends on how large an oscillation
 amplitude will be found in experiment: If the azimuthal variation remains 
 within the limited range found in our calculations even in a fragmentation-dominated
 momentum range, its discriminating 
 power will be low, due to statistical uncertainties; a much larger 
 oscillation signal could, however, eliminate all of the models studied 
 in this work.}.
We confirm earlier findings that it is not easy to reproduce
the relatively large oscillation amplitude of $R_{AA}(\phi)$ found by
the PHENIX experiment. We identified several mechanisms that help to
increase the in-plane vs. out-of-plane spread of $R_{AA}$ and, in 
combination, may be able to explain the data: (i) A stronger path length
dependence of parton energy loss, combined with (ii) a delayed beginning
of the energy loss action, caused by the need for allowing the scattering
centers in the medium that induce the hard parton to radiate energy to
decohere from the initial state wave function of the colliding nuclei,
(iii) a delay of the flow-induced dilution of the medium density
by viscous heating, and (iv) a larger fireball eccentricity by intializing
the hydrodynamic evolution with CGC-fKLN initial conditions rather than the
less deformed Glauber model initial density profile. We found that the
combination of effects (i) and (ii) accounts for about 50\% of the 
azimuthal spread found in our calculations, viscous heating contributes 
another 35\% of the effect, with the remaining 15\% arising from different 
initial fireball eccentricities.

Analyzing the reasons why the mechanisms (i)-(iii) cause a larger 
in-plane vs. out-of-plane spread of $R_{AA}$, we found that they all
shift the weight of the energy loss towards later times along the path 
of the parton. This allows the parton energy loss to probe the spatial
anisotropy of the medium (which is a global fireball property) on a 
global length scale, rather than probing the properties of the medium
locally in the vicinity of the production vertex only. If the global 
spatial anisotropy of the medium can be probed by a parton, the 
connection between spatial medium anisotropy and final state hard 
parton momentum anisotropy is strongest.

We would like to point specifically to the role of viscosity in this context:
We are beginning to see significant roles played by the (necessarily 
non-zero) shear viscosity of the quark-gluon plasma even though its 
absolute value (expressed through the dimensionless ratio $\frac{\eta}{s}
={\cal O}(1{-}3)\times\frac{1}{4\pi}$) is almost as small as theoretically 
possible \cite{Policastro:2001yc}. In \cite{SHHS} we found that 
$\frac{\eta}{s} \approx (2{-}3)\times\frac{1}{4\pi}$ gives the best fit 
to the pion and proton momentum spectra in $200\,A$\,GeV Au-Au collisions,
and here we found that a similar value contributes significantly to the
in-plane vs. out-of-plane spread of $R_{AA}$ for pions. We hope and 
expect that a combined analysis of all RHIC data on soft hadron
production {\em and} hard parton medium modification will eventually lead to 
an accurate determination of the quark-gluon plasma viscosity.  

\begin{acknowledgments}
Discussions with Kari Eskola are gratefully acknowledged. This work was 
supported by an Academy Research Fellowship of T.R. from the Finnish Academy (Project 130472) and 
from Academy Project 133005. H.H. gratefully acknowledges the financial
support from the national Graduate School of Particle and Nuclear Physics.
The research of U.H. and C.S. was supported  by the U.S.\ Department of
Energy under contract DE-SC0004286 and within  the framework of the JET
Collaboration under grant number DE-SC0004104.

\end{acknowledgments}

\end{document}